# The Calibrated Deepfake Trust Score (CDTS): Competence-Coupled Trust Degradation Across Deepfake Detectors

Md Anas Biswas
*School of Computing, University of Portsmouth, Portsmouth, United Kingdom*
up2082724@myport.ac.uk · ORCID: 0009-0009-0113-5816

## Abstract

In moderation, provenance, and verification pipelines a deepfake detector's output probability is read as a degree of trust, so its calibration matters as much as raw accuracy. We reframe deepfake detection as a calibrated, self-auditing trust instrument, the Calibrated Deepfake Trust Score (CDTS), and identify what governs its trustworthiness. Our central finding is a competence–trust coupling with a sharp division: the raw score's miscalibration tracks discriminative competence almost perfectly ($r = -0.98, -0.98, -0.95$ across two convolutional networks and a CLIP vision transformer), and a calibrator deployed without target labels fails with competence just as tightly ($r = -0.98$ on the primary detector, for isotonic, Platt, and beta calibrators alike). Given target labels, by contrast, any well-specified calibrator repairs any detector, including inverted ones: in-domain calibratability is not competence-limited, and the trust failure is a distribution-shift phenomenon concentrated exactly on the low-competence generators that motivate deployment. Explanation faithfulness rises and falls on the same competence axis. We reach this conclusion after uncovering, and correcting, a tie-handling degeneracy in the standard equal-mass expected-calibration-error (ECE) estimator that fabricates strong spurious competence–calibration coupling on tie-heavy calibrated scores; the same degeneracy invalidates calibration-equity gaps we previously reported, a caution for fairness auditing. Competence is trackable without labels: batch predictive entropy flags generators with high deployed calibration error at ROC-AUC 0.99, and routing source-batches on label-free competence beats confidence-based routing precisely in the low-competence regimes the coupling identifies, while confidence regains the advantage where competence is high. Trust scoring must be competence-aware; CDTS is the mechanism.

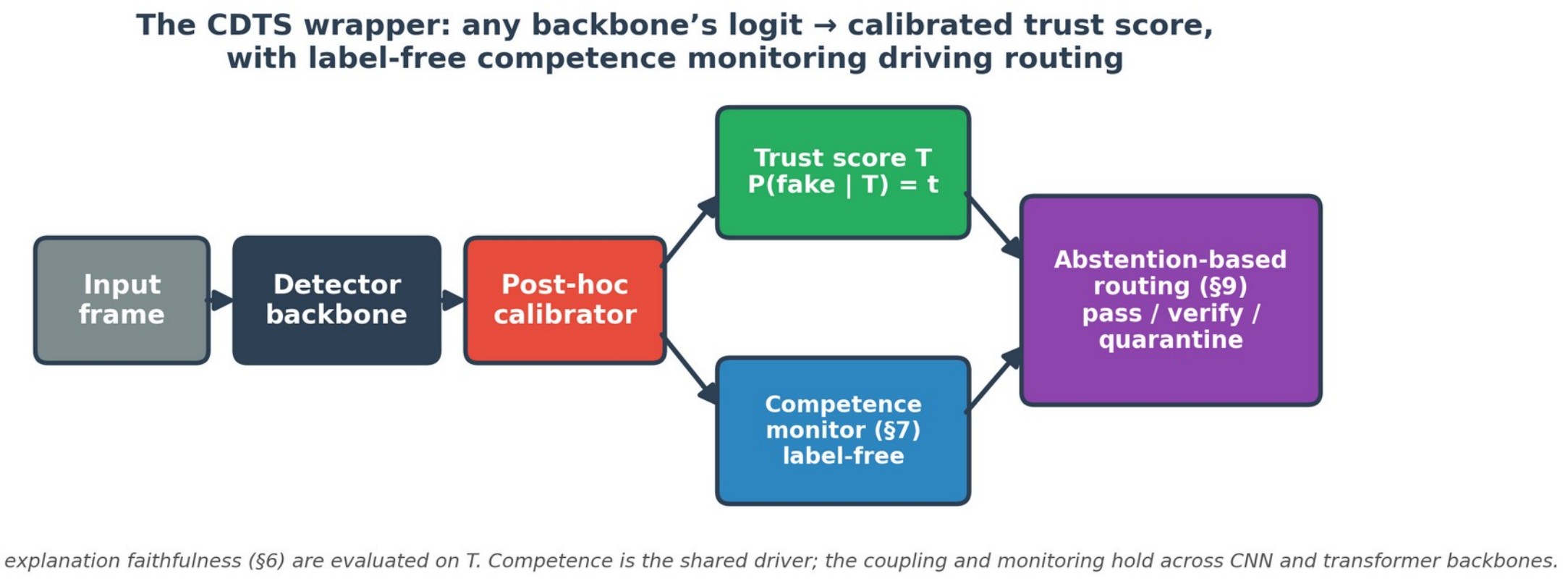

**Figure 1.** *The CDTS wrapper. Any backbone's logit is mapped through a post-hoc calibrator to a trust score, augmented with a label-free competence monitor that drives an abstention-based routing policy. The equity audit and explanation-faithfulness analysis are evaluated on the trust score. Detection competence is the shared driver of these trust properties; the coupling and label-free monitoring hold across CNN and transformer backbones (Sections 4, 7.2).*

# 1 Introduction

## 1.1 Detectors are consumed as trust signals

Deepfake detection is no longer an academic exercise in binary classification. As synthetic media has moved into political communication, fraud, and non-consensual imagery, detectors have been embedded into operational pipelines, content moderation at platform scale, media-provenance frameworks, identity verification, and journalistic verification, where their output is not a verdict but an input to a downstream decision. In these settings a detector's probability is interpreted as a degree of trust: a moderation system escalates content in proportion to how confident the detector is that it is manipulated, and a verification workflow passes or holds an asset based on where its score falls. What is consumed is therefore the calibrated meaning of the probability, not merely the sign of a thresholded decision.

This shift makes calibration a first-order property. A detector whose 0.9 outputs are wrong nine times in ten is dangerous in a way that a detector with the same accuracy but honest uncertainty is not, because confident errors propagate silently: a confidently mislabelled asset is never escalated for review. The reliability of the probability, formalized as calibration (Guo et al., 2017), is thus as consequential as discriminative accuracy for any system that reads the score as trust. Yet the deepfake-detection literature overwhelmingly optimizes and reports accuracy (AUC, detection rate), and the small body of work that treats the probability itself (Jin et al., 2025) does so as a single-axis calibration-improvement problem rather than asking how calibration behaves as the detector encounters the novel generators that motivate the field.

## 1.2 Graceful Trust Degradation and the CDTS instrument

We adopt a simple organizing principle for what a trust instrument owes its consumer: it must never fail silently. As a detector's reliability degrades, under a new generator, heavy compression, or for particular demographic subgroups, the trust signal should **know** it (remain calibrated, or report that it cannot), **show** it (offer explanations whose faithfulness is itself diagnostic), **warn** about it (detect the degradation, ideally without labels), **not discriminate** in it (degrade evenly across subgroups), and **route** around it (feed a verification policy that abstains when trust is low). We call this Graceful Trust Degradation (GTD).

The Calibrated Deepfake Trust Score (CDTS) is the artifact that operationalizes GTD (Figure 1). CDTS is a wrapper, not a new backbone: it maps any detector's logit through a post-hoc calibrator to a calibrated trust score T, a calibrated fake-probability with target $P(\text{fake} \mid T = t) = t$ (an authenticity score is simply $1 - T$), and augments it with a label-free competence estimate and a routing policy. Treating the instrument, rather than the detector, as the unit of contribution lets us ask a question the accuracy literature cannot: when these five trust properties degrade, do they degrade independently, each requiring its own guarantee, or together, governed by something common?

## 1.3 From a temporal conjecture to competence coupling

Our initial hypothesis, and the one around which this project was originally designed, was temporal: that as generators evolve over time, calibration drift, explanation instability, and fairness degradation would fire together, and that a drift alarm could forecast their joint collapse before accuracy visibly dropped. The experiments reported here revise that hypothesis in an important way. We find **no independent temporal effect**: once detection competence is controlled for, the apparent timeline degradation vanishes (the partial correlation of generator release era with calibration error, holding competence fixed, is approximately zero and non-significant; Section 7). The variable that actually unifies the trust signals is not the calendar but **competence**, how well the detector discriminates on the generator at hand. Time matters only as a proxy for how far a generator sits from the detector's competence frontier. This is a stronger and more actionable claim than the temporal conjecture, and we rebuild the paper around it, reporting the revision openly because it is itself a finding.

This paper also corrects its own earlier account. A preliminary version reported that calibration remains poor even when a calibrator is fitted on the target generator's own labels, a claim we traced to a tie-handling degeneracy in the equal-mass ECE estimator (Section 3.5) that fabricates calibration error on tie-heavy calibrated scores. Under a corrected estimator the picture inverts in-domain and sharpens in deployment: with target labels, calibration is cheap at any competence; without them, it fails with competence at $r = -0.98$. We document the correction in full, both because the corrected result is stronger and because the failure mode, beyond an order- and tie-break sensitivity previously noted for regression-uncertainty binning (Pernot, 2023), appears undocumented in the deterministic, label-ordered form that produced our artifact, and likely affects other published calibration pipelines.

## 1.4 Contributions

**C1. The competence–trust coupling, and where it lives (Section 4).** The raw trust score's miscalibration tracks discriminative competence at $r = -0.98, -0.98, -0.95$ across three architecturally distinct detectors, and, on the primary detector, a single calibrator frozen on in-domain data fails with competence at $r = -0.98$ when deployed to unseen generators, for isotonic, Platt, and beta calibrators alike ($r = -0.96$ to $-0.98$ on the strictly out-of-domain generators alone). Given target labels, by contrast, any well-specified calibrator achieves ECE between 0.008 and 0.06 across all 61 suite configurations (AUC 0.43–0.99), and the orientation audit of Appendix C extends the same conclusion to the four induced-low-competence configurations down to AUC 0.26, including detectors with inverted scores: calibratability in-domain is not what competence buys; the trust failure lives in deployment under shift.

**C2. A tie-handling degeneracy in equal-mass ECE (Section 3.5).** Equal-mass binning implemented with a stable sort assigns tied predictions to bins in input row order; on label-ordered data a constant-prevalence predictor, which is calibrated up to prevalence shift, is charged ECE on the order of $2p(1-p)$ instead of the prevalence shift. Because isotonic calibration collapses toward a constant precisely on low-signal and inverted-signal data, the artifact selectively inflates low-competence calibrated ECE and fabricates a spurious competence–calibration coupling; it fabricated the target-fitted coupling and the calibration-equity gaps of our own preliminary report. We provide a corrected estimator, a regression test, and a full orientation-and-calibrator audit.

**C3. Calibration-equity auditing is vulnerable to the same degeneracy (Section 5).** Under the corrected estimator, subgroup calibration gaps that were previously significant vanish: every worst-versus-pooled and pairwise gap interval spans zero at 70 identities, and a competence-adjusted cell-level regression finds no independent subgroup effect. We report the null and the audit protocol as a caution for the growing calibration-fairness literature.

**C4. Explanation faithfulness is competence-coupled (Section 6).** Under a saturation-robust per-region metric, Grad-CAM faithfulness rises with competence ($r = +0.94$); this analysis involves no ECE and is unaffected by the correction.

**C5. Label-free monitoring of deployable calibration risk (Section 7).** Predictive entropy tracks competence without labels with a consistent negative sign on all three backbones and flags generators with high deployed (transferred) calibration error at ROC-AUC 0.99 (95% CI [0.94, 1.00]); reference-divergence signals reverse sign across architectures and are not portable.

**C6. Competence as the shared driver, formally tested (Section 8).** On the deployable quantity, the trust signals share one competence-aligned structure: transferred ECE correlates with competence at $-0.99$, with faithfulness at $-0.91$, and with the monitoring signals at $|r| \geq 0.83$, and a paired-difference bootstrap shows every pairwise association is significantly attenuated by conditioning on competence (all $\Delta$ intervals exclude zero).

**C7. Competence-aware routing, evaluated across architectures (Section 9).** Routing source-batches on label-free competence beats confidence-based routing on both convolutional pools, where competence is low and confidence is untrustworthy, while per-frame confidence retains the advantage on the high-competence transformer pool; a reference-divergence router is actively harmful under the sign reversal of C5. The regime in which each signal wins is itself predicted by the coupling.

Throughout, we report uncertainty on the claims that bear on deployment: the raw and transferred couplings carry generator-level and family-clustered bootstrap intervals, every equity statement carries identity-clustered pivotal intervals, mediation is tested by paired-difference bootstrap rather than asserted from attenuation, and we document each confound we broke, each deviation from our original design, and the estimator failure we corrected.

## 2 Related Work

Five literatures bear on this paper, each studying in isolation a property that we connect to a single factor. We position our contribution against the closest prior work in each, and against the one existing attempt to calibrate deepfake detectors specifically.

### 2.1 Deepfake detection and benchmarks

Detector architectures range from naive CNN backbones to frequency-aware and foundation-model detectors; we build on two canonical reproducible baselines, Xception (Chollet, 2017) and EfficientNet-B4 (Tan & Le, 2019), as implemented in the DeepfakeBench framework (Yan et al., 2023). Our empirical timeline and cross-generator analysis use FaceForensics++ (Rössler et al., 2019), Celeb-DF v2 (Li et al., 2020), and the DF40 generator suite (Yan et al., 2024), which spans GAN-era to diffusion-era methods and supplies a controlled progression of forgery techniques. Our contribution, however, is orthogonal to the choice of backbone: CDTS wraps a detector's score, so the trust properties we study are properties of the wrapper and the score, not of any one architecture. The broader evidence that detectors age, that in-the-wild performance collapses against contemporary generators (Chandra et al., 2025), is precisely the deployment reality that makes calibration-under-novelty, rather than calibration on a fixed benchmark, the question worth asking.

### 2.2 Calibration

Post-hoc calibration is mature for general classifiers: Platt scaling fits a sigmoid to the scores (Platt, 1999), isotonic regression fits a non-parametric monotone map (Zadrozny & Elkan, 2002), temperature scaling provides a single-parameter softmax rescaling (Guo et al., 2017), and beta calibration fits a two-parameter family whose orientation is unconstrained (Kull et al., 2017). Calibration is quantified by the Expected Calibration Error (ECE) and its relatives; because ECE is sensitive to its binning scheme, debiased and equal-mass estimators are preferred (Kumar et al., 2019; Roelofs et al., 2022), a caution we adopt by reporting equal-mass ECE throughout. The single most relevant prior work is Jin et al. (2025), which introduces deepfake-detection calibration via packed ensembles and reports reliability diagrams and ECE on standard benchmarks. That work treats calibration as a quantity to improve on a fixed test distribution. It does not study how calibration behaves as competence varies across generators, does not address subgroup calibration, explanation faithfulness, or monitoring, and does not connect any of these. Our competence–calibration coupling reframes the problem: the question is not only how to calibrate, but where calibration can be trusted, and the answer is governed by competence. We adopt the bootstrap-CI rigor of Jin et al. (2025) as our competitive floor. That calibration degrades under dataset shift is well established for general classifiers (Ovadia et al., 2019; Minderer et al., 2021); Section 4.3 gives that phenomenon a governing variable in this domain and a near-perfect coefficient. A complementary calibrated-abstention line certifies authenticity rather than calibrating a detector: Hashmi et al. (2025) replace the binary real-versus-fake verdict with a calibrated, inversion-based authenticity index that abstains on plausibly-deniable content. That framework changes what is scored; CDTS calibrates and audits the score an existing detector already emits, and the two are compatible layers of one verification pipeline. Concurrent with this revision, Postiglione, Gortner, and Subrahmanian (2026) compare 200 human raters with 95 automated detectors and report, via reliability diagrams on DF40 among other data, that detector confidence is poorly calibrated and carries weak diagnostic information about correctness. That study documents the miscalibration descriptively across a detector population; our contribution identifies the variable that governs it, competence, separates the repairable in-domain regime from the transfer regime in which deployment actually fails, and supplies the label-free monitoring and routing that act on the diagnosis. On estimation, the nearest prior observation to the failure mode we document in Section 3.5 is Pernot (2023), who shows that isotonic recalibration stratifies regression uncertainties into ties whose bin assignment, and hence bin-based statistics such as the ENCE, becomes sensitive to data order and to the tie-breaking of the sorting algorithm, an aleatoric fluctuation. Section 3.5 shows the classification-ECE analogue is worse than fluctuation: with a stable sort on label-ordered data the bias is deterministic rather than random, maximal for collapsed calibrators, and selective for weak models, strong enough to fabricate a quality–calibration coupling and subgroup calibration gaps.

## 2.3 Fairness in deepfake detection

Fairness work in this domain is framed almost entirely as accuracy disparity. Gender-balanced benchmarks (Nadimpalli & Rattani, 2022), massively-annotated demographic databases (Xu et al., 2024), and million-scale demographically-annotated face datasets with multi-detector fairness benchmarks (Lin et al., 2025) all measure error-rate or AUC gaps across demographic groups. Closest to our setting in spirit is recent label-free fairness work: Face-Fairness (Brown & Russell, 2026) is a detector-agnostic, demographic-label-free framework that remaps a detector's logits, conditioned on frozen face embeddings, to shrink the across-group false-positive and true-positive gaps without access to identity attributes. Our aim is different along two axes. First, the target quantity: Face-Fairness equalizes an accuracy-style criterion (group error and FPR/TPR gaps), whereas we measure and audit calibration equity, whether the trust score's probability remains reliable within each subgroup, a conceptually distinct target from accuracy parity; Section 5 reports our audit of it, a bounded null under a corrected estimator, together with a caution for how it is measured. Second, the role of the remapping: their logit remapping is an intervention that improves fairness, whereas our calibrator is a diagnostic instrument: its transferred residual error exposes where competence is too low for the deployed score to be trustworthy at all, while its target-fitted error shows what labels could repair. The two are complementary: a label-free remapper like Face-Fairness could sit upstream of the calibration-equity audit we propose. Our calibration-equity measurement is grounded in multicalibration (Hébert-Johnson et al., 2018), which requires simultaneous calibration across identifiable subgroups rather than only on average. Multicalibration makes this conceptually distinct from accuracy fairness: a subgroup could in principle be classified equally well yet carry a worse-calibrated trust score. Section 5 builds the audit for exactly that question and reports a bounded null under the corrected estimator. We use the demographic annotations of Xu et al. (2024) for this analysis.

## 2.4 Explanation faithfulness and selective prediction

Explanation faithfulness, whether a saliency map identifies the regions a model actually uses, is measured by deletion/insertion curves and area-over-perturbation metrics (Samek et al., 2017; Petsiuk et al., 2018; DeYoung et al., 2020), typically over gradient-based saliency such as Grad-CAM (Selvaraju et al., 2017), whose model-dependence is itself subject to sanity checks (Adebayo et al., 2018). The one deepfake-specific evaluation of explanation faithfulness (Gowrisankar & Thing, 2024), argues that generic removal-based evaluation is ill-suited to deepfake detectors and proposes an adversarial alternative. We show in Section 6 that the standard deletion metric is confounded for our purpose by score saturation, and introduce a saturation-robust per-region rank metric; we then use it not to compare explainers but to expose that faithfulness itself rises and falls with competence. For routing, we draw on selective prediction and the reject option (El-Yaniv & Wiener, 2010; Geifman & El-Yaniv, 2017), reporting risk-coverage curves; the algorithmic-recourse direction (Mothilal et al., 2020) we leave to future work. The label-free monitoring view in Section 7 borrows the activity-monitoring framing of detecting interesting change without labels (Fawcett & Provost, 1999), but applies it to competence rather than to a class signal.

### 2.5 What is new here

Each property above has been studied alone. The contribution of this paper is to show they are not separate: calibration, calibration equity, explanation faithfulness, and label-free monitorability are governed by a common factor, competence, which is the right object to estimate and condition on. We establish this across three architectures spanning convolutional networks and a vision transformer, and we characterize its scope: the raw coupling and entropy-based label-free monitoring generalize across all three; the transferred coupling is established on the primary backbone, calibrator-agnostic there, with cross-backbone transfer left to future work; the faithfulness analysis covers both convolutional backbones; and the in-domain coupling vanishes under correct measurement. To our knowledge this unification, its mediation-consistent structure formally tested by paired-difference bootstrap, the cross-architecture scoping, the estimator correction of Section 3.5, and the resulting competence-aware routing are new to deepfake detection, as is the calibration-equity audit protocol itself.

## 3 The CDTS Instrument and Experimental Setup

### 3.1 The CDTS wrapper and notation

Let a detector produce a logit for an input frame, mapped by a softmax to a raw fake-probability p (the model's estimated probability that the frame is manipulated). CDTS applies a post-hoc calibrator g, fit on a held-out identity-disjoint calibration split, to obtain the calibrated trust score $T = g(p)$. We adopt the convention that T is a calibrated fake-probability: the calibration target is $P(\text{fake} \mid T = t) = t$, so that $T = 0.9$ should correspond to a 90% empirical frequency of manipulation, and all calibration error (ECE) is computed on T against the is-fake label. A consumer who prefers an authenticity score uses $1 - T$; we report and calibrate the fake-probability direction throughout because it is the detector's native output and the exact quantity whose calibration we measure (equal-mass ECE is not perfectly invariant under the $1 - T$ inversion, so we fix the direction here to make the reported quantity unambiguous). The explanation-faithfulness analysis (Section 6) perturbs this same fake-probability. We measure discriminative *competence* by the area under the ROC curve (AUC) on a given generator, *calibration* by the Expected Calibration Error of the calibrated score (ECE_cal), and downstream trust properties on T itself. The wrapper additionally maintains a label-free competence estimate from the score distribution (Section 7), which drives the routing policy (Section 9). Nothing in the construction is backbone-specific.

### 3.2 Detectors and the competence-manipulation control

We use three detectors from the DeepfakeBench framework (Yan et al., 2023), all with published weights and all trained on face-swap (FS) forgeries, which keeps the cross-architecture comparison fair. The primary detector is Xception (Chollet, 2017). To test architecture-generality within the convolutional family we add EfficientNet-B4 (Tan & Le, 2019). To test whether our findings extend beyond convolutional networks entirely, we add a third detector built on a CLIP vision transformer (Radford et al., 2021), a non-convolutional foundation-model backbone (the DeepfakeBench CLIP detector, fine-tuned with low-rank adapters on the same FS training data and loaded from its published checkpoint); it is scored at its native 224x224 resolution against the 256x256 of the two CNNs. The CLIP detector exposes a pooled embedding rather than a spatial convolutional feature map, so it admits no Grad-CAM saliency; it therefore enters the competence-calibration coupling (Section 4) and the label-free monitoring analysis (Section 7.2) but not the explanation-faithfulness or intrinsic-factor analyses, which require a per-region saliency map (Sections 6, 8.2). To break the confound between competence and dataset identity (Section 4.4), we additionally use a reenactment-trained (FR) Xception checkpoint, which is broadly incompetent on face-swap content (all AUC < 0.67); applying it to DF40 generators induces low competence on otherwise-detectable forgeries, manipulating competence while holding the images fixed. Grad-CAM (Selvaraju et al., 2017) is computed at the final convolutional feature map of each convolutional backbone.

### 3.3 Data

Four sources are used overall: three for the main analyses and the FaceForensics DeepFakeDetection set (DFD) as an external-consistency source. FaceForensics++ c23 (Rössler et al., 2019) supplies a 22,388-frame test split, real frames included, across four forgery methods (faceswap, deepfakes, face2face, neuraltextures); pooled FF++ calibration numbers are reported on its 11,190-frame identity-disjoint evaluation half. The split carries the per-frame demographic annotations of Xu et al. (2024) (gender, age band, ethnicity, and a skin-appearance attribute (shiny skin)), joined to our scored frames by identity. Celeb-DF v2 (Li et al., 2020) supplies a 16,420-frame test split and the real frames paired with the DF40 generators. DF40 (Yan et al., 2024) supplies the generator suite for the timeline, monitoring, faithfulness, and unifying analyses; we use its Celeb-DF-sourced face-swap, reenactment, and entire-face-synthesis methods. Following the dataset's identity structure, demographic labels transfer by identity for the face-swap and reenactment subsets built on FaceForensics++/Celeb-DF identities; entire-face-synthesis methods synthesize new identities and are therefore excluded from the subgroup-calibration cells. DF40 is used under its CC BY-NC 4.0 (non-commercial) license. For the external-validity test of Section 4.5 we additionally use DFD, a fourth source the detectors never trained on, from which we score 2,493 face crops across 40 manipulated and 40 real videos.

### 3.4 Calibration, metrics, and statistics

We calibrate per target generator, not once in-domain, and every protocol detail is fixed as follows. For each generator we split its scored frames into two identity-disjoint halves (50/50, seed 42), fit the calibrator on one half using that generator's labels, and report calibration error on the held-out half; the identity-disjoint split prevents leakage between the calibration and evaluation halves. The default calibrator is hybrid, selected by calibration-set size: Platt scaling (Platt, 1999) when the calibration set is small ($n < 1000$, where a low-variance parametric fit is preferable) and isotonic regression (Zadrozny & Elkan, 2002) otherwise; the calibrator grid of Section 4.2 additionally reports unconstrained Platt and beta calibration (Kull et al., 2017) across the full competence range, and the transferred setting of Section 4.3 is shown to be calibrator-agnostic. We report AUC for competence and ECE for calibration, computed with 15 equal-mass bins under the tie-safe estimator of Section 3.5 throughout (equal-mass binning is less biased than equal-width (Roelofs et al., 2022), and where the standard estimator is at risk we report the debiased view (Kumar et al., 2019)); values under the uncorrected legacy estimator appear only where explicitly marked (Section 3.5). For the sharpness claim of Section 4.2 we additionally report the bin-based reliability–resolution decomposition of the Brier score (Murphy, 1973) under the same binning.

Per-generator calibrated measurement uses a target-fitted protocol: for each generator, a fresh calibrator is fitted on an identity-disjoint calibration half of that generator's own frames and evaluated on the held-out half. The reported ECE is therefore the best case achievable for that generator if its labels were available, subject to finite-sample fitting error, split choice, and calibrator specification; we do not treat it as a strict lower bound, and Section 4.3 reports the one generator (rddm) on which the transferred error undercuts it. We use the target-fitted protocol because it isolates the question we study in Section 4.2, what competence permits when labels exist, and we contrast it throughout with the raw score and with the transferred setting of Section 4.3, which is the quantity a deployment actually consumes.

We report bootstrap 95% confidence intervals on the central couplings and on every interval-based claim: generator-level intervals (the coupling coefficients and the mediation paired differences of Section 8) use $B = 5000$ resamples with the percentile interval, and the subgroup-equity intervals use $B = 2000$ identity-clustered resamples with the pivotal interval; correlations elsewhere are reported with significance tests rather than intervals. For the subgroup analysis, bootstraps resample whole identities (clustered) and use the pivotal interval, because frames nest within identities and a naive frame-level bootstrap is anti-conservative.

Correlations are reported as Pearson with Spearman alongside; partial correlations control for AUC by linear residualization; the dependence-aware family-cluster and leave-one-family-out analyses appear in Section 4.7. Appendix B inventories every configuration set used in the paper and the sections in which each appears.

### 3.5 The ECE estimator: a tie-handling degeneracy and its correction

Equal-mass ECE partitions predictions into equal-count bins by sorting them and splitting the sorted order. When the sort is stable, tied predictions are assigned to bins in input row order. For continuous scores this is irrelevant, but calibrated scores are often tie-heavy: isotonic regression outputs piecewise-constant values, and on data where the score carries weak or inverted signal it collapses toward a single constant near the prevalence. In that regime the estimator becomes degenerate. Write $p_{cal}$ for the prevalence of the calibration split, which is the value a collapsed calibrator predicts as a constant, and $p_{eval}$ for the prevalence of the evaluation split. Treating the tied predictions as the single tie block they are, the empirical calibration error of the constant predictor is $|p_{eval} - p_{cal}|$, the prevalence shift, which is what a correct estimator must charge. Under stable-sort binning of label-ordered rows, however, the bins are label-pure blocks: a fraction $1 - p_{eval}$ of bins are pure-real, each charged $|p_{cal} - 0| = p_{cal}$, and a fraction $p_{eval}$ are pure-fake, each charged $|p_{cal} - 1| = 1 - p_{cal}$, for a total of $(1 - p_{eval})\cdot p_{cal} + p_{eval}\cdot(1 - p_{cal})$, which reduces to $2p(1 - p)$ when the two prevalences coincide. The observed instance is exact: on the DF40 evaluation splits, $p_{cal} = 0.838$ and $p_{eval} = 0.830$, so the correct charge is $|0.830 - 0.838| = 0.008$, the ideal label-pure value is $0.170\cdot 0.838 + 0.830\cdot 0.162 = 0.277$, and the measured legacy value is 0.268, the small shortfall from the ideal reflecting one mixed boundary bin.

The consequence for any study relating calibration to model quality is severe, because the degeneracy is selective: it fires exactly where the model is weak, where isotonic collapses, and therefore manufactures a negative quality–ECE association out of nothing (Pernot (2023) notes the order- and tie-sensitivity of binning stratified calibrated outputs in regression-uncertainty statistics; the label-ordered stable-sort case here is its deterministic, selective extreme). We correct the estimator by breaking ties with an infinitesimal seeded perturbation before sorting, so tied predictions are distributed across bins uniformly at random and the estimator recovers the correct conditional-mean behaviour; we verified agreement with an exact estimator that groups predictions by unique value. A regression test, a constant-prevalence predictor over label-ordered rows must score approximately the prevalence shift, is committed with the code, and the uncorrected estimator is retained under an explicit legacy name so that every number in the preliminary version of this work remains reproducible. All ECE values in this paper use the corrected estimator; where we quote a value from the preliminary version for contrast, we mark it as legacy.

## 4 The Competence–Trust Coupling

We begin with the finding the rest of the paper rests on, stated with the division our own correction (Section 3.5) makes visible: the coupling is real and near-perfect, but it lives in the raw score and in deployment under shift, not in what labels can repair. Every ECE below uses the corrected estimator; values quoted from the preliminary version for contrast are marked legacy.

### 4.1 Raw trust scores degrade with competence

We first measure the quantity every pipeline starts from: the detector's own probability, uncalibrated. Across the full generator suites of all three detectors, 61 configurations spanning AUC 0.43 to 0.99, raw ECE rises as competence falls at $r = -0.982$ (Xception, 21 generators), $-0.983$ (EfficientNet-B4, 20), and $-0.950$ (CLIP, 20), with raw error ranging from 0.02 on the most competent configurations to 0.58 on the least (Figure 3, left; Table 2). Because the evaluation prevalence is high (0.83 fake), we verify in Section 4.7 that the coupling is not a prevalence or global-bias artifact: it survives a shift-only bias correction and re-evaluation at balanced prevalence. The relationship is near-deterministic and architecture-independent: an uncalibrated deepfake detector is overconfident in almost exact proportion to its incompetence. This is the first face of the coupling, and because raw scores are continuous it is untouched by the estimator issue of Section 3.5.

## 4.2 Given target labels, calibration is not competence-limited

The preliminary version of this work reported the opposite of what we now establish: that even a calibrator fitted on the target generator's own labels leaves low-competence configurations badly miscalibrated. Under the corrected estimator, and with attention to calibrator specification, that claim is false. Three points establish it. First, the theoretical floor: at any competence, the constant-prevalence forecast is calibrated up to prevalence shift, so calibration alone is always attainable; what a detector at AUC 0.5 cannot provide is sharpness, the resolution term of the classical reliability–resolution decomposition (Murphy, 1973). We measure both terms rather than asserting them: decomposing the target-fitted score's Brier score into reliability and resolution with the same tie-safe equal-mass bins, the reliability term is small and competence-flat on every detector (0.0002–0.0054; r against AUC +0.024, +0.270, and −0.343, all non-significant), while the resolution term rises with competence at r = +0.835, +0.921, and +0.923 (pooled +0.868 across the 61 configurations, all $p < 3 \times 10^{-6}$): competence buys resolution, and only resolution. Second, orientation: an AUC below 0.5 signals an inverted, not absent, ranking, which an increasing-only isotonic map cannot exploit but an unconstrained logistic (Platt) or beta calibrator (Kull et al., 2017) inverts automatically, their fitted slope simply turning negative; an orientation audit (Appendix C) confirms the fitted Platt slope is negative on all four induced, genuinely inverted configurations, while the near-0.5 EfficientNet configurations show slopes near zero of either sign, as expected when ranking signal is absent rather than inverted. Third, and empirically decisive: with the corrected estimator, target-fitted calibration achieves ECE between 0.008 and 0.060 across all 61 suite configurations (the 21 Xception, 20 EfficientNet-B4, and 20 CLIP generator configurations of Section 4.1, AUC 0.43–0.99) and all three calibrators tested (hybrid, Platt, beta), with no significant dependence on competence on Xception (r = +0.01) or EfficientNet (r = +0.39, a weak positive), and a modest residual negative dependence only on CLIP (r = −0.51, 95% CI [−0.79, −0.10]), whose continuous score distribution never triggered the estimator degeneracy and whose residual coupling we therefore report as the one surviving in-domain effect. The four induced-low-competence configurations of Section 4.4 (AUC 0.26–0.46) are a distinct set, not part of the 61; the orientation audit of Appendix C shows Platt and beta calibration achieve ECE 0.02–0.07 on them as well, so the conclusion extends to the full induced range. Restricting to the deployment-relevant band of AUC 0.6–0.85, the pooled target-fitted coupling is r = −0.08 to −0.19 and non-significant under every calibrator. Table 4 reports the full calibrator-by-estimator grid, including the legacy values this paper corrects. Cheap is meant literally, not only as attainability under the 50% target-fitted split: subsampling the calibration set shows that roughly 100 labeled frames suffice for beta calibration to reach median ECE 0.03–0.05 at every competence we test, including the inverted AUC-0.26 configuration, with the 90th percentile over 25 resamples at or below 0.08 and label cost flat in competence. In-domain, with labels, calibration is cheap at any competence.

**Table 1.** *Representative configurations under the corrected estimator: raw, target-fitted best case (beta calibrator), and transferred (frozen in-domain calibrator) ECE, FS-Xception, plus the pooled reenactment-checkpoint FF++ row; the transferred setting applies to the DF40 suite only. Legacy values in parentheses are the preliminary version's target-fitted figures under the degenerate estimator. Target-fitted calibration succeeds at every competence; the raw and transferred errors carry the coupling.*

| Generator | AUC | Raw ECE | Target-fitted ECE (legacy) | Transferred ECE |
|---|---|---|---|---|
| blendface | 0.943 | 0.081 | 0.017 (0.022) | 0.040 |
| ddim | 0.766 | 0.197 | 0.018 (0.041) | 0.180 |
| DiT | 0.521 | 0.475 | 0.015 (0.161) | 0.432 |
| pixart | 0.433 | 0.580 | 0.024 (0.265) | 0.514 |
| Xception-FR pooled FF++ | 0.547 | 0.694 | 0.019 (0.207) | n/a |

## 4.3 The coupling lives in deployment: a frozen calibrator fails with competence

Deployment never has target labels. We fit a single calibrator once, on a pooled identity-disjoint calibration split of the high-competence in-domain generators (simswap, blendface, facedancer, fsgan, faceswap, inswap), freeze it, and apply it unchanged to every generator's held-out evaluation frames. The transferred ECE rises as competence falls at $r = -0.979$ across all 21 generators and $r = -0.964$ on the 15 strictly out-of-domain generators alone, and the result is calibrator-agnostic: isotonic, Platt, and beta transfer give $r = -0.979$, $-0.983$, and $-0.979$ respectively ($-0.964$, $-0.975$, and $-0.964$ on the out-of-domain generators alone), with per-generator errors agreeing to within 0.01 (Figure 5; Table 2). On the high-competence generators the frozen calibrator is nearly as good as the target-fitted best case (mean ECE 0.033 for AUC $> 0.85$); on the low-competence generators it collapses (mean ECE 0.451 for AUC $< 0.60$, a fourteen-fold increase). One generator, rddm, exceeds the pattern (transferred ECE 0.303 at AUC 0.733), the single exception to the target-fitted-below-transferred ordering, which we attribute to its anomalous score distribution (Section 7). This is the deepfake instance of the general finding that calibration fails under dataset shift (Ovadia et al., 2019), here with the size of the failure governed by a single measurable variable. Together with Section 4.1, this locates the coupling precisely: competence governs how overconfident the raw score is and how badly a deployed calibrator fails under shift, not whether the score is calibratable in principle.

**Table 2.** *The coupling across data slices under the corrected estimator: Pearson correlation of AUC against raw ECE per slice, and against transferred ECE for the deployable setting. All slices use tie-safe equal-mass ECE; identity-disjoint evaluation halves throughout.*

| Slice (quantity) | n | Pearson r | AUC range |
|---|---|---|---|
| FF++ Xception, 2 ckpt × 4 methods (raw) | 8 | −0.949 | 0.39–0.97 |
| DF40 within, FS + FR-mismatch (raw) | 12 | −0.934 | 0.26–0.95 |
| Xception full suite (raw) | 21 | −0.982 | 0.43–0.95 |
| EfficientNet-B4 full suite (raw) | 20 | −0.983 | 0.47–0.94 |
| CLIP-ViT full suite (raw) | 20 | −0.950 | 0.56–0.99 |
| Grand raw pool, all slices | 77 | −0.904 | 0.26–0.99 |
| Xception transferred, all generators | 21 | −0.979 | 0.43–0.95 |
| Xception transferred, out-of-domain only | 15 | −0.964 | 0.43–0.82 |

## 4.4 Breaking the confound by inducing low competence

A correlation between competence and trust degradation across generators could be confounded by generator identity. We break it by manipulating competence while holding images fixed: applying the reenactment-trained Xception checkpoint to four DF40 methods spanning face-swap and entire-face synthesis (simswap, blendface, StyleGAN2, sd2.1) forces AUC to 0.26–0.46 on frames the face-swap-trained checkpoint scores at 0.65–0.95. On the raw quantity the within-DF40 coupling across the 8 face-swap-trained and 4 mismatch configurations is $r = -0.934$ (Figure 2). Because the images are fixed and only the checkpoint changed, dataset and generator identity are eliminated as explanations of the degradation; the checkpoint swap, however, moves competence together with everything else that distinguishes the two training regimes, so we read the manipulation as removing the data-identity confound rather than as isolating competence as the sole cause, and the causal support rests on the monotone within-DF40 relationship across the induced range.

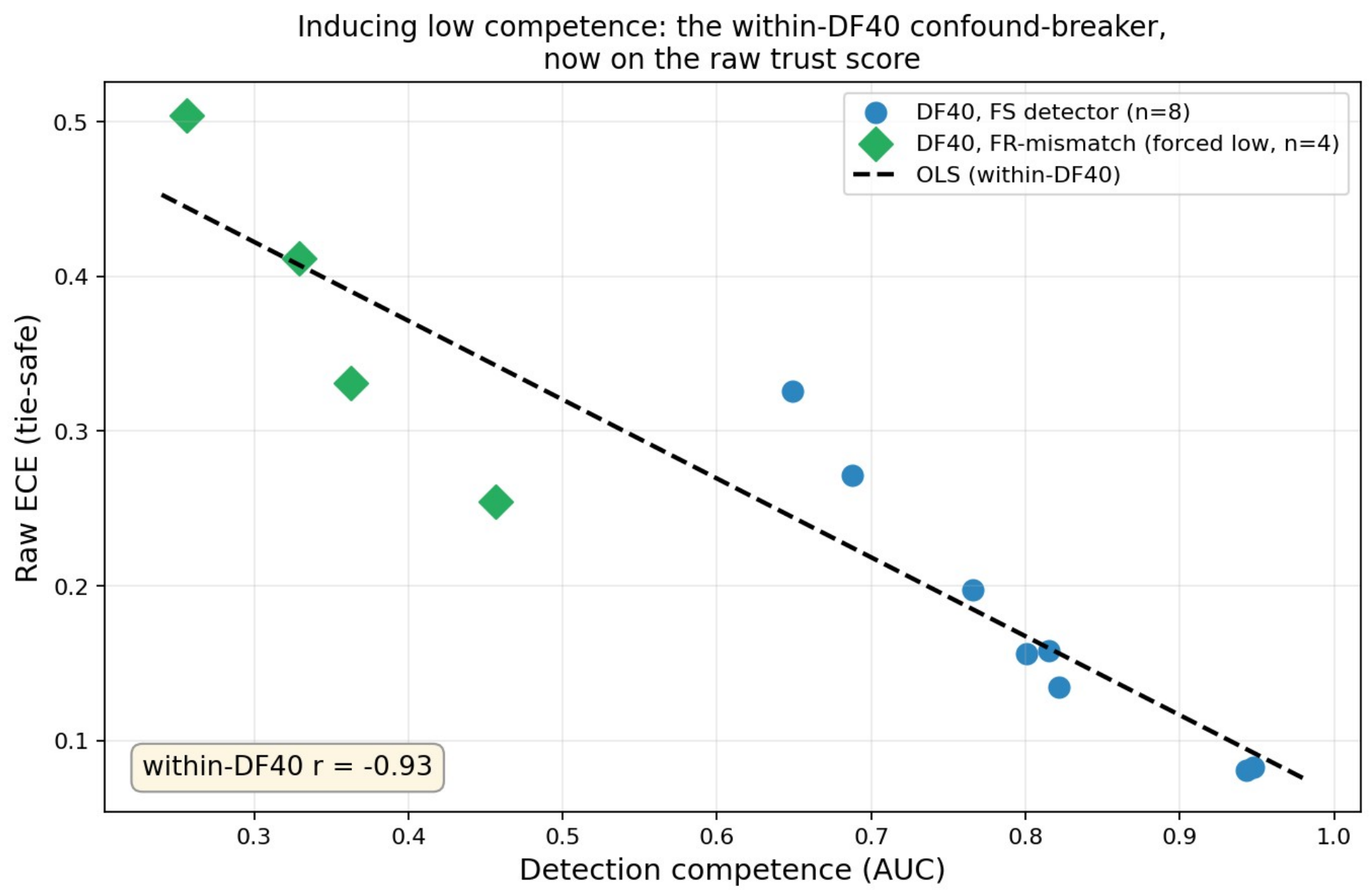

**Figure 2.** *Inducing low competence: the within-DF40 confound-breaker on the raw trust score. Applying a mismatched (reenactment-trained) checkpoint craters AUC and inflates raw ECE on otherwise-detectable forgeries, removing generator identity as the explanation for the degradation (r = −0.93, tie-safe estimator).*

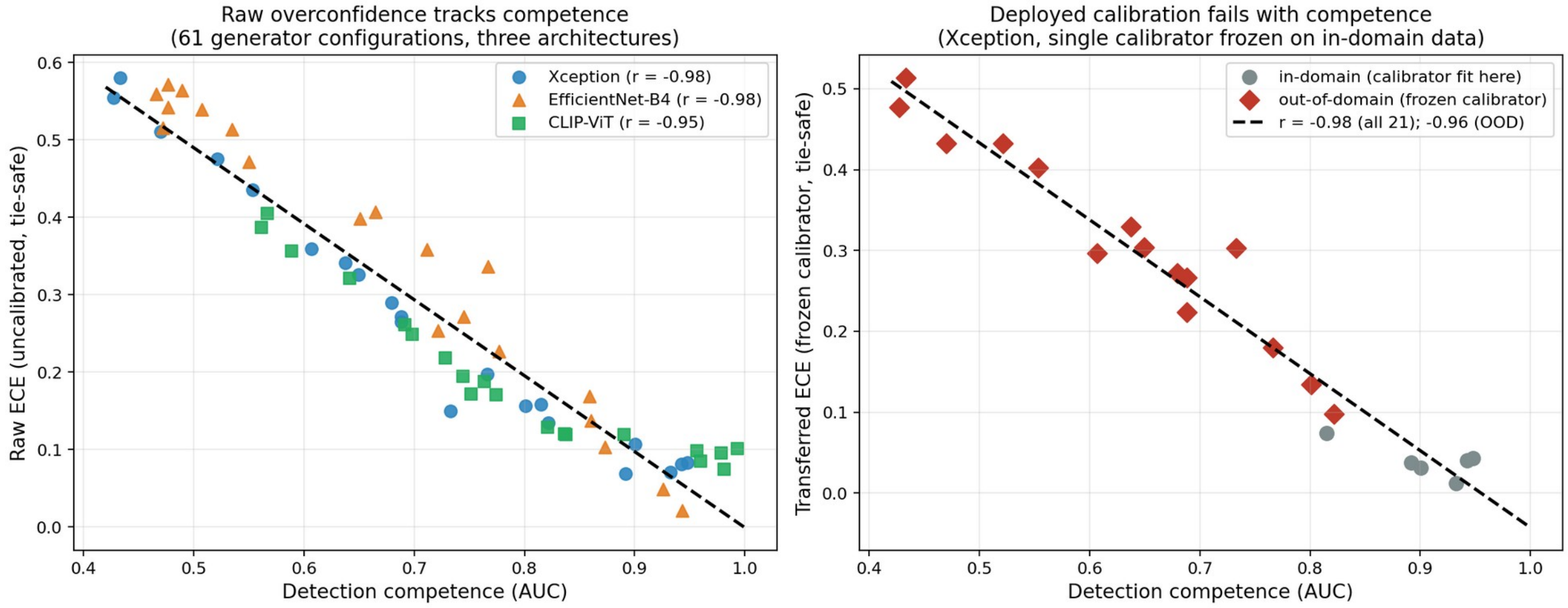

**Figure 3.** *The competence–trust coupling under the corrected estimator. Left: raw ECE against competence across the 61 suite configurations of all three architectures (per-detector r = −0.98, −0.98, −0.95). Right: transferred ECE for a single calibrator frozen on in-domain data and deployed to all 21 Xception generators (r = −0.98; out-of-domain generators marked). Given target labels, by contrast, calibration succeeds everywhere (Table 4).*

## 4.5 External consistency on a fourth dataset (DFD)

To check the coupling on data the detectors never trained on and that we did not use to fit anything, we score the FaceForensics DeepFakeDetection set (DFD), 2,493 face crops across 40 manipulated and 40 real videos, with all three FS-trained detectors. Two views agree with the coupling. First, the whole-dataset raw ECEs (0.285 Xception, 0.265 EfficientNet, 0.219 CLIP, on identity-disjoint evaluation halves) sit against the 61-configuration raw regression as Figure 4 shows: two of the three points fall inside the 95% prediction interval and the third (EfficientNet) falls below it, better calibrated than predicted, so no detector exceeds the predicted risk. Second, within DFD, per-actor competence against per-actor raw calibration error over the 16 actors with both manipulated and real frames gives a strong negative coupling on every detector (Table 3), r = −0.89 (Xception), −0.77 (EfficientNet), and −0.63 (the transformer), all significant at $p < 0.01$, over a wide per-actor range (AUC 0.14 to 1.00 across detectors; 0.17 to 1.00 on Xception itself). We state the scope plainly: with three detector points, one manipulation family, and sixteen actors, this is an external consistency check, not strong external validation, and the per-actor analysis uses raw ECE, which is the artifact-free quantity throughout.

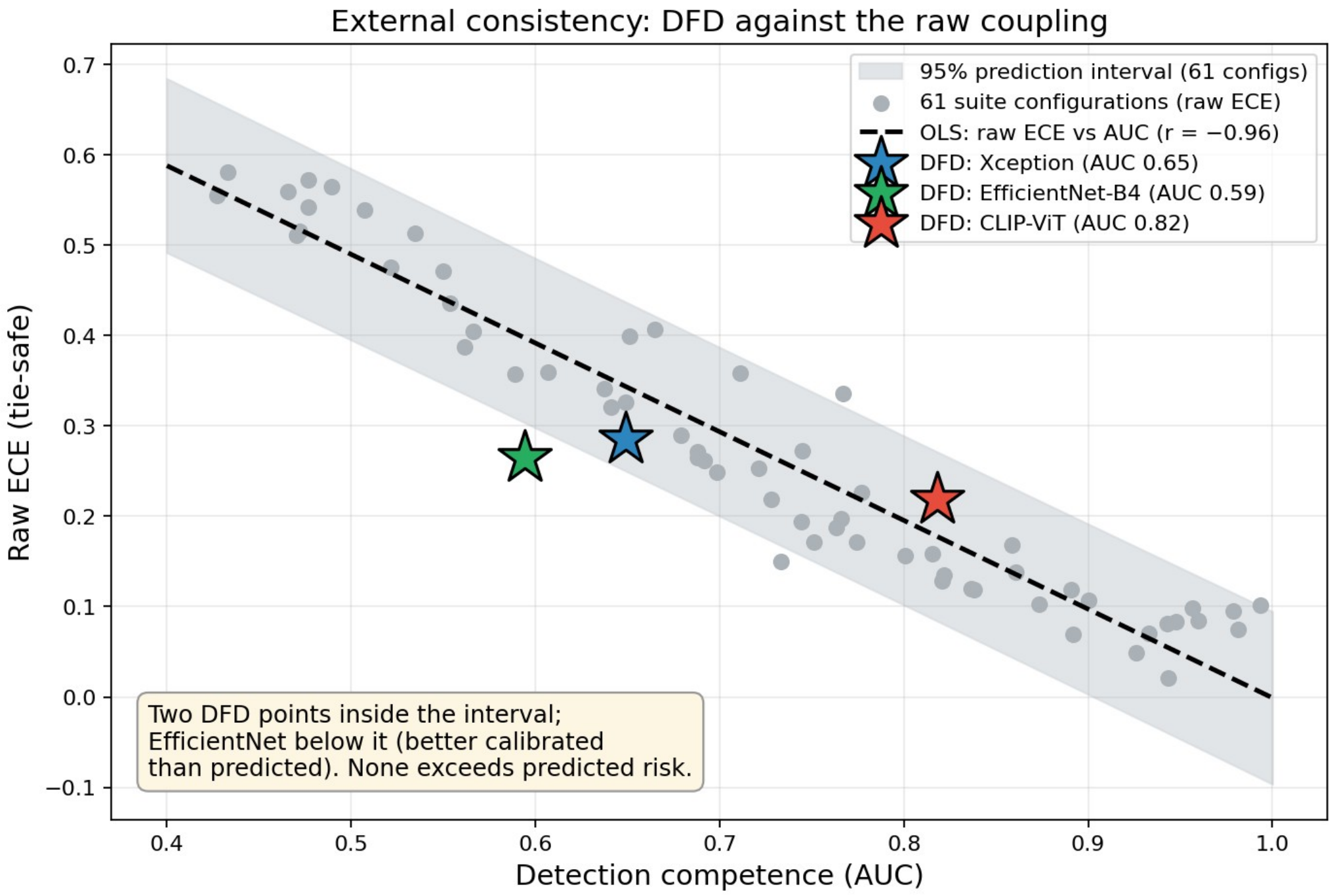

**Figure 4.** *External consistency on a fourth dataset. The three DFD raw-ECE points (stars) against the 61-configuration raw coupling with its 95% prediction interval (shaded). Two points fall inside the interval and EfficientNet falls below it; none exceeds the predicted risk.*

**Table 3.** *Within-DFD per-actor competence–calibration coupling on a fourth, held-out dataset (FaceForensics DeepFakeDetection). Per-actor Pearson and Spearman correlation of competence (AUC) against calibration error, over the 16 actors with both manipulated and real frames, for each FS-trained detector. Per-actor ECE is the raw (not re-calibrated) error of that small slice. The coupling holds on all three architectures over a wide per-actor competence range (AUC 0.14–1.00).*

| Detector | Per-actor AUC range | Pearson r | Spearman | p |
|---|---|---|---|---|
| Xception | 0.17–1.00 | −0.89 | −0.86 | <0.001 |
| EfficientNet-B4 | 0.14–1.00 | −0.77 | −0.76 | <0.001 |
| CLIP-ViT | 0.24–1.00 | −0.63 | −0.81 | 0.009 |

## 4.6 Why this happens

A calibrator learns the score-to-frequency map on the distribution it is fitted on. Given target labels, that map exists at any competence, down to the constant-prevalence map at AUC 0.5, which is why in-domain calibration succeeds everywhere; competence buys sharpness, not calibratability, and Section 4.2 measures both halves of that decomposition. Deployed under shift, the frozen map is evaluated on a score distribution it never saw, and the size of that mismatch is governed by how far the generator sits from the detector's competence frontier, which is why the transferred error tracks competence at $r = -0.98$. The consequence for deployment is unchanged from the preliminary version and now rests on sound measurement: trust scores are least reliable exactly where competence is lowest, which is exactly the novel generators a detector has not seen, the case that matters most (Chandra et al., 2025).

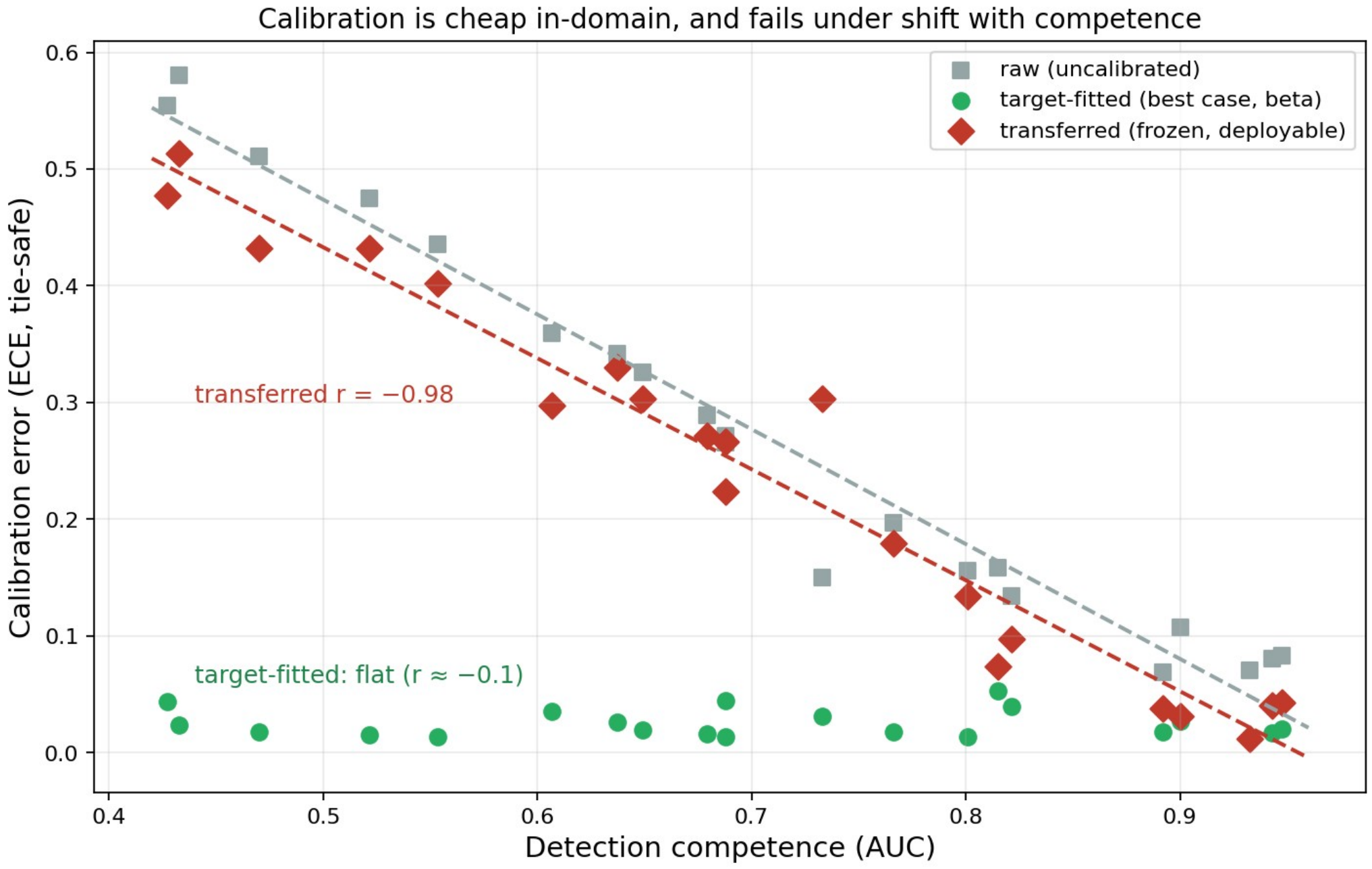

**Figure 5.** *Calibration is cheap in-domain and fails under shift with competence. Raw (squares), target-fitted best case (beta calibrator, green), and transferred (frozen, deployable; diamonds) ECE against competence on Xception. The target-fitted error is flat; the raw and transferred errors carry the coupling.*

**Table 4.** *The correction in one table: coupling coefficient r (AUC against ECE) per detector under the legacy and corrected estimators and per calibrator. The legacy estimator manufactures a strong negative coupling for the target-fitted setting on all three detectors; under the corrected estimator the target-fitted coupling vanishes on the CNNs and only a modest residual survives on CLIP, while the raw coupling is near-perfect everywhere.*

| Detector | Raw | Target-fitted (legacy est.) | Target-fitted (corrected) | Platt | Beta |
|---|---|---|---|---|---|
| Xception | −0.982 | −0.864 | +0.008 | −0.020 | −0.094 |
| EfficientNet-B4 | −0.983 | −0.833 | +0.290 | +0.126 | +0.393 |
| CLIP-ViT | −0.950 | −0.859 | −0.542 | −0.427 | −0.509 |

## 4.7 Robustness and dependence

The raw and transferred couplings are stable across bin counts, rank-correlation alternatives, split seeds, and, for the transferred setting, calibrator choice. Excluding all AUC $< 0.5$ configurations, substituting orientation-free $\text{AUC}^* = \max(\text{AUC}, 1 - \text{AUC})$, and restricting to the deployment band leave both couplings strong (raw: −0.939, −0.943, and −0.847 in the 0.6–0.85 band, $n = 29$, $p = 7 \times 10^{-9}$; transferred: −0.969, −0.959, and −0.902 in-band, $n = 11$), while the same manipulations reduce the target-fitted coupling to zero, confirming the division of Sections 4.2–4.3 (Table 5). Because configurations share architectures, datasets, checkpoints, and forgery families, we treat pooled p-values as descriptive and additionally report dependence-aware inference: a cluster bootstrap over forgery families gives a raw-coupling 95% interval of [−0.978, −0.810] and a transferred interval of [−0.999, −0.724], and leaving out any one forgery family changes the raw coefficient by at most 0.005 and the transferred coefficient by at most 0.015. Both AUC and ECE are estimated quantities; at the evaluation-split sizes here (roughly 10,000 to 17,000 frames per generator) their sampling error is an order of magnitude below the effects reported. A separate control addresses global probability bias and class prevalence. At the DF40 evaluation prevalence of 0.83, a large share of raw ECE is global miscentering: the bias component |mean score − prevalence| accounts for a median 78% of raw ECE and itself tracks competence ($r = −0.92, −0.98, −0.80$ per detector), which is the decomposed content of the raw coupling, global overconfidence growing with incompetence. The coupling is not, however, reducible to that component or to the prevalence choice: after a shift-only correction fitted on the calibration half (intercept only, slope fixed at one), the residual shape miscalibration still couples at r = −0.93, −0.95, and −0.97, with band means 0.05 (AUC above 0.85) against 0.21 (AUC below 0.60); and re-evaluating raw ECE on class-balanced (50/50) subsamples of each evaluation half gives r = −0.95, −0.99, and −0.85. Both the bias term and the shape residual are competence-coupled, and neither the prevalence nor a single global offset explains the coupling away.

**Table 5.** *Robustness of the corrected couplings: exclusion of AUC < 0.5, orientation-free AUC*, the deployment band, family-cluster bootstrap, and leave-one-family-out. The target-fitted row is the contrast: the same manipulations that leave the raw and transferred couplings intact reduce it to zero.*

| Quantity | Full | Excl. AUC<0.5 | AUC* | Band 0.6–0.85 |
|---|---|---|---|---|
| Raw (pooled 61) | −0.962 | −0.939 | −0.943 | −0.847 (p=7e−9) |
| Transferred (21) | −0.979 | −0.969 | −0.959 | −0.902 (p=1e−4) |
| Target-fitted (pooled 61) | −0.08 to −0.19 | n.s. | n.s. | n.s. |

## 5 Calibration Equity: a Null Result and a Caution

If competence governs calibration, and competence varies, then calibration should vary across any partition of the data on which the detector's competence varies, including demographic subgroups. We therefore ask whether the trust score remains calibrated within subgroups, not merely on average.

### 5.1 The subgroup calibration gap

Using the per-frame demographic annotations of Xu et al. (2024) joined by identity to the FS-Xception trust scores on the FaceForensics++ test split (the equity audit is single-detector; Section 10.3), we adopt a multicalibration view (Hébert-Johnson et al., 2018), computing every ECE below with the corrected estimator of Section 3.5: a trust score should be calibrated simultaneously across identifiable subgroups. We summarize per-axis disparity by the worst-vs-pooled subgroup-ECE gap, the worst subgroup's ECE minus the pooled ECE computed over all frames, with identity-clustered pivotal bootstrap intervals (B = 2000); clustering is necessary because frames nest within identities and the evaluation split contains only 70 identities, so a frame-level bootstrap would badly understate uncertainty. This worst-vs-pooled gap is the multicalibration quantity; it is distinct from the pairwise subgroup contrast reported in Table 7 (Section 5.2), and the two can differ when the worst subgroup is neither of the two groups being contrasted.

### 5.2 Previously reported gaps do not survive the corrected estimator

The preliminary version of this work reported significant calibration-equity gaps on the gender, age, and skin-appearance axes. Those gaps were computed with the degenerate estimator of Section 3.5 on isotonic-calibrated scores, exactly the tie-heavy regime in which it is degenerate, and they do not survive correction. Under the corrected estimator every per-subgroup ECE falls by a factor of two to four (the worst cell, middle-aged, from 0.115 to 0.066; Figure 6, left), every worst-versus-pooled gap interval spans zero (Table 6), and every pairwise contrast interval spans zero on both the accuracy and the calibration axis, with the gender calibration gap changing sign (Table 7). A competence-adjusted cell-level analysis (36 subgroup-by-forgery-method cells; ECE regressed on cell AUC and axis indicators with identity-clustered bootstrap over 70 identities) finds no independent subgroup association on any axis; the only significant coefficient is a positive association between cell AUC and cell ECE (+0.27, 95% CI [+0.09, +0.30]), consistent with the in-domain picture of Section 4.2 in which target-fitted calibration error is not driven down by competence. At 70 identities we therefore report no detectable calibration-equity disparity for this detector. We label the fourth axis a skin-appearance attribute deliberately: the underlying annotation is shiny skin, a photometric appearance attribute in Xu et al. (2024), not a demographic skin-tone scale such as Fitzpatrick or Monk, and we do not treat it as one.

**Table 6.** *Worst-vs-pooled subgroup-ECE gap under the corrected (tie-safe) estimator, with identity-clustered pivotal 95% bootstrap confidence intervals (B = 2000, 70 identities). Every interval spans zero: no calibration-equity disparity is detectable at this sample size.*

| Axis | Worst-vs-pooled gap | 95% CI (clustered pivotal) | CI excludes 0 |
|---|---|---|---|
| gender | +0.012 | [−0.006, +0.019] | no |
| age | +0.045 | [−0.018, +0.062] | no |
| ethnicity | +0.030 | [−0.096, +0.053] | no |
| skin appearance | +0.011 | [−0.015, +0.018] | no |

## 5.3 What the audit is worth

The null is informative in two directions. Methodologically, it is a caution: calibration-fairness auditing inherits every pathology of the calibration estimator, and the tie degeneracy is worst precisely for small subgroups scored by collapsed calibrators, so equity gaps computed with stable-sort equal-mass ECE on calibrated scores should be treated as suspect until re-estimated. Substantively, the multicalibration question stands and the audit protocol we release, identity-clustered pivotal intervals, pairwise and worst-versus-pooled statistics, and competence adjustment, is the right instrument for it; 70 identities simply lacks the power to detect gaps of the size that survive correction, and we bound rather than claim. Accuracy-gap fairness work (Nadimpalli & Rattani, 2022; Lin et al., 2025) is unaffected by this issue, since AUC involves no binning; the caution applies specifically to the calibration-equity extension we and others have begun to pursue.

**Table 7.** *Pairwise subgroup contrast under the corrected estimator: signed accuracy gap (AUC, computed on the trust score) and signed calibration gap (ECE) between the two named subgroups per axis, with identity-clustered pivotal 95% intervals (B = 2000). Every interval on both axes spans zero; the preliminary version's significant calibration gaps were an artifact of the estimator degeneracy of Section 3.5.*

| Axis (group 1 vs group 2) | AUC gap, 95% CI | ECE gap, 95% CI | Either excludes 0? |
|---|---|---|---|
| gender (M vs F) | −0.006 [−0.048, +0.032] | −0.013 [−0.039, +0.002] | no |
| skin appearance (shiny vs not) | +0.003 [−0.039, +0.047] | +0.005 [−0.012, +0.037] | no |
| age (young vs senior) | +0.049 [−0.007, +0.136] | −0.010 [−0.026, +0.052] | no |

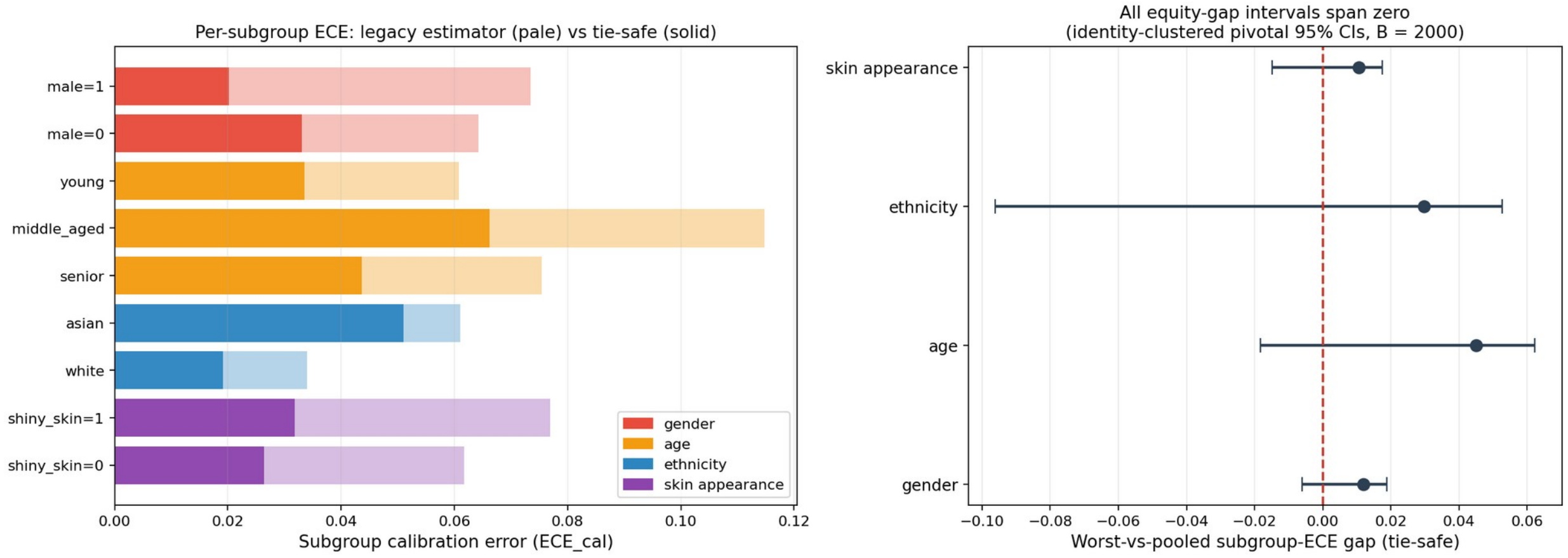

**Figure 6.** *Left: per-subgroup calibration error under the legacy estimator (pale) and the corrected tie-safe estimator (solid); every subgroup's measured error falls by a factor of two to four. Right: worst-vs-pooled gap intervals under the corrected estimator; all span zero.*

# 6 Explanation Faithfulness

A trust instrument that ships explanations should ship faithful ones, and the faithfulness of those explanations is itself a trust signal. We show that explanation faithfulness, properly measured, is governed by the same competence factor as calibration.

## 6.1 A saturation-robust faithfulness metric

Standard deletion/insertion faithfulness (Petsiuk et al., 2018; DeYoung et al., 2020) is confounded for our purpose by score saturation. A competent detector pins the fake-probability near 1.0 (the start-score correlates with competence at $r = 0.998$), which compresses the dynamic range of any absolute-score perturbation curve and can even reverse its apparent sign; we observed this directly when an initial deletion-curve analysis produced a competence ranking opposite to every other metric. We therefore use a scale-free measure. We partition each image into an 8×8 grid; for each patch we record its Grad-CAM (Selvaraju et al., 2017) importance and the actual change in fake-probability when only that patch is removed; faithfulness is the Spearman rank correlation between importance and effect across the 64 patches. This asks whether the regions the explanation emphasizes are the regions that actually move the decision, a question whose answer is independent of the absolute score level, so the saturation confound cannot arise by construction.

## 6.2 Faithfulness rises with competence

Across 20 generators, explanation faithfulness increases with competence: Pearson $r = +0.94$, $p < 10^{-7}$ (Figure 7). A competent detector's saliency identifies regions that genuinely drive its decision; an incompetent detector's saliency is uncorrelated with the evidence the model actually uses, an explanation in form only. Figure 8 makes the mechanism concrete at the per-region level: for competent detectors (simswap, blendface) Grad-CAM importance predicts the real per-region effect, producing a rising trend, whereas for incompetent detectors (pixart, wav2lip) importance and effect are unrelated, producing a flat scatter. In these experiments, the same explanation machinery is faithful or not largely according to detector competence.

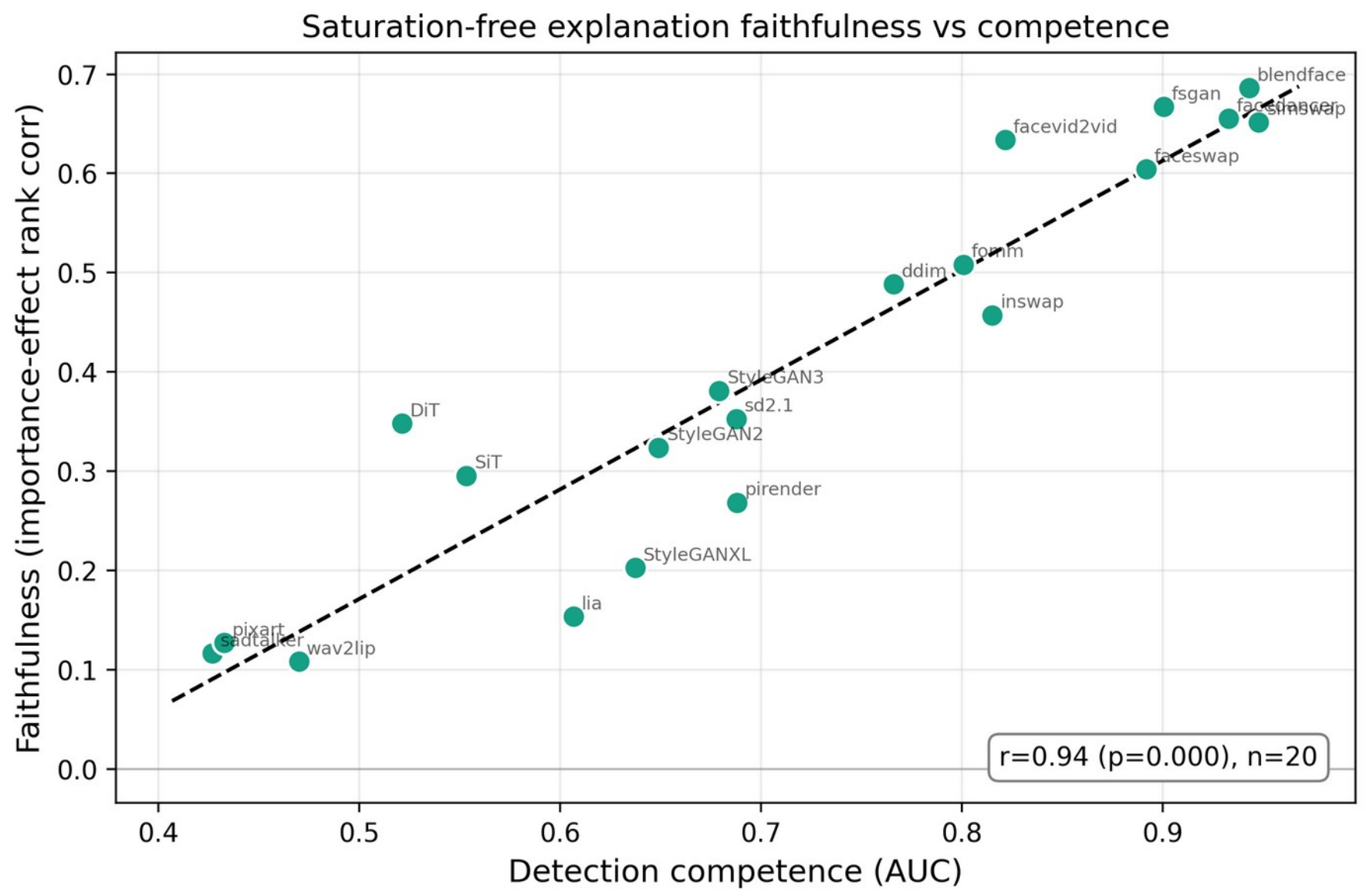

**Figure 7.** *Saturation-free explanation faithfulness against competence across 20 generators (r = +0.94). When a detector cannot discriminate, its Grad-CAM explanation carries no signal that our per-region probe can detect.*

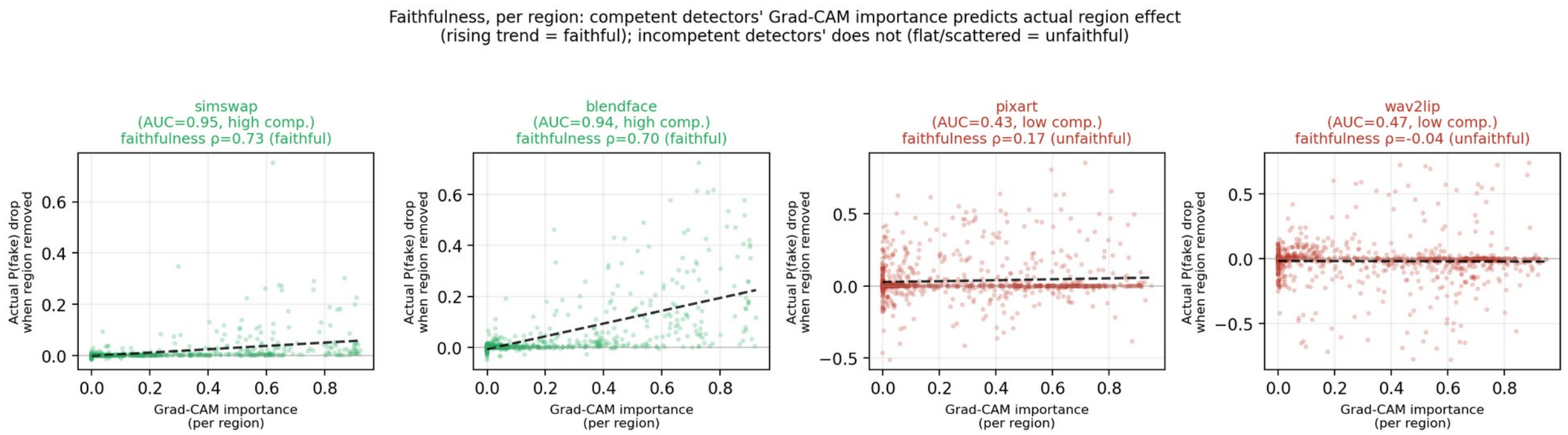

**Figure 8.** *Per-region faithfulness. For competent detectors (left two panels), Grad-CAM importance predicts the actual fake-probability drop when a region is removed (rising trend, faithful); for incompetent detectors (right two), importance is uncorrelated with effect (flat, unfaithful). The metric is saturation-robust by construction.*

# 7 Label-Free Competence Monitoring

The coupling is only deployable if competence can be estimated when it matters, on a fresh generator for which no labels exist. This section shows it can, and in doing so corrects our original temporal hypothesis.

## 7.1 There is no independent temporal drift

We scored 21 DF40 generators spanning release eras 2016–2024 (Table 8). Newer generators are harder (era against competence $r = -0.30$) and their deployed calibration is worse (era against transferred ECE $r = +0.36$), which might suggest a temporal-drift story. It is not one, and the corrected quantities make the point more cleanly than the preliminary version could: raw miscalibration correlates with era at only +0.30 (non-significant at $n = 21$), and controlling for competence reduces it to +0.01; the transferred partial is +0.36 and likewise non-significant. Within the face-swap family, where the detector is competent, transferred ECE stays between 0.01 and 0.07 across eras spanning 2016 to 2023. Whatever a generator's release date, the trust score degrades when, and only insofar as, competence drops; release era carries no significant independent signal. This corrects our original temporal conjecture: the drift that motivated this study is competence drift.

## 7.2 Two families of label-free signal, and which one generalizes

In deployment AUC is unavailable because labels are. But an unlabeled signal that tracks competence inherits, through the coupling, the ability to forecast calibration risk. Label-free signals computed from the predicted-probability stream fall into two families, and the distinction matters because they do not generalize equally across architectures. The first family is reference-divergence: the distance of a generator's score distribution from an in-domain reference distribution (Kolmogorov–Smirnov, Wasserstein). On Xception these track competence strongly (KS vs AUC $r = 0.961$ on the 20-generator unifying set (the timeline minus rddm, scored by all three detectors; Appendix B); 0.952 on the 21-generator timeline set, Table 9). The second family is reference-free uncertainty: statistics of the detector's own output that need no reference, namely predictive entropy and score dispersion (the standard deviation of scores). On Xception these also track competence (entropy vs AUC $r = -0.837$ on the 20-generator unifying set; Table 9 reports the same signals on the 21-generator timeline set). The gap between the two sets is one generator: rddm, a late addition to the timeline set, is the most influential point for the timeline correlations, and leave-one-out re-estimation moves entropy from −0.68 to −0.84 and dispersion from −0.74 to −0.81 when it is removed; every monitoring correlation keeps its sign under leave-one-out on both sets.

On a single architecture the two families look interchangeable, but they are not. Testing the same signals on EfficientNet-B4 and on the CLIP transformer separates them sharply (Table 10). The reference-divergence signal decouples from competence on EfficientNet (KS vs AUC $r = -0.339$, from +0.961 on Xception):

distance-from-reference depends on the backbone's score geometry, not competence alone, and is therefore not a portable monitor. The decisive difference between the families is directional consistency, not raw magnitude. Predictive entropy never reverses sign: it correlates negatively with competence on all three backbones (Xception $r = -0.837$, EfficientNet $r = -0.384$, CLIP $r = -0.951$), and score dispersion is strongly negative on the two architectures where the score distribution has range (Xception $r = -0.739$ on the 21-generator timeline set, CLIP $r = -0.962$); on EfficientNet the dispersion compresses to a narrow band (standard deviation 0.29 to 0.37 across the 20 generators) and its residual correlation with competence is weakly positive ($r = +0.53$, 95% CI [+0.17, +0.81]), the opposite sign to the other two backbones, so dispersion is not a dependable cross-architecture monitor. Reference-divergence is worse: KS correlates with competence at +0.961 on Xception but reverses to −0.339 on EfficientNet and −0.94 on the CLIP transformer (against an in-domain reference; Table 10), and Wasserstein likewise flips from +0.90 to −0.09 and −0.97. Two of the three backbones thus reverse the Xception orientation, so the anomaly is Xception's score geometry, not a CNN-versus-transformer divide.

This sign reversal is the failure that matters operationally. A monitor is a threshold on a signal; if the signal-to-competence relationship is strongly positive on one backbone and negative on another, a threshold tuned on the first backbone flags exactly the wrong generators on the second. A reference-free signal that is weak but correctly signed everywhere is a usable monitor; a reference-divergence signal that is strong but sign-unstable across architectures is not. EfficientNet is the hard case for all distribution-shape signals, reference-free included: its score geometry compresses these statistics, so entropy is only −0.384 there, weaker than on the other two backbones though still correctly signed. The practical conclusion is therefore narrower and more defensible than a blanket magnitude claim: label-free competence monitoring should be built on predictive entropy, whose direction is stable across CNN and transformer backbones (and on score dispersion where the score distribution has range), rather than on reference-divergence, whose sign is not, and a deployment should expect the signal to be stronger on some backbones than others.

Through the coupling, a label-free monitor now targets the quantity that matters in deployment: the transferred calibration error of Section 4.3. Fourteen of the 21 timeline generators exceed transferred ECE 0.1, and batch predictive entropy separates them from the safe generators at ROC-AUC 0.990 (generator-level bootstrap 95% CI [0.94, 1.00]); KS divergence reaches 0.94 on this backbone but is not portable for the sign-reversal reason above (Figure 9). We state the claim at the strength the evidence supports: batch predictive entropy carries strong label-free information about competence and deployable calibration risk, with a direction that is consistent across all three backbones; it is not yet a calibrated competence estimator with prediction intervals and family-held-out validation, which we leave to future work. The reference-divergence comparison carries one further caveat: the CLIP reference is an in-domain DF40 face-swap pool while the CNN references are FaceForensics++, so the cross-architecture sign reversal mixes an architecture change with a reference change; scoring a common reference pool with all three detectors is required for a fully clean portability statement, though the EfficientNet reversal alone, measured against the same reference as Xception, already establishes non-portability. We note one signal, the mean predicted score, correlates even more strongly with competence but is confounded by class balance and is excluded as a deployable signal (Table 9); its information content is exactly the global-bias component made explicit in Section 4.7.

**Table 8.** *The 21-generator timeline ordered by release era, with competence and the two artifact-free trust quantities: raw ECE and transferred ECE (frozen in-domain calibrator, corrected estimator). Family codes: FS face-swap, FR face reenactment, EFS entire-face synthesis. Newer eras are harder and worse-transferred, but era carries no significant signal once competence is controlled (Section 7.1).*

| Generator | Era | Family | AUC | Raw ECE | Transferred ECE |
|---|---|---|---|---|---|
| faceswap | 2016 | FS | 0.892 | 0.069 | 0.037 |
| fomm | 2019 | FR | 0.801 | 0.156 | 0.134 |
| fsgan | 2019 | FS | 0.900 | 0.107 | 0.031 |
| wav2lip | 2019 | FR | 0.470 | 0.511 | 0.432 |

| Generator | Era | Family | AUC | Raw ECE | Transferred ECE |
|---|---|---|---|---|---|
| StyleGAN2 | 2020 | EFS | 0.649 | 0.326 | 0.304 |
| ddim | 2020 | EFS | 0.766 | 0.197 | 0.180 |
| simswap | 2020 | FS | 0.948 | 0.083 | 0.043 |
| StyleGAN3 | 2021 | EFS | 0.679 | 0.289 | 0.271 |
| facevid2vid | 2021 | FR | 0.822 | 0.134 | 0.097 |
| pirender | 2021 | FR | 0.688 | 0.265 | 0.223 |
| DiT | 2022 | EFS | 0.521 | 0.475 | 0.432 |
| StyleGANXL | 2022 | EFS | 0.637 | 0.341 | 0.330 |
| lia | 2022 | FR | 0.607 | 0.360 | 0.297 |
| sd2.1 | 2022 | EFS | 0.688 | 0.272 | 0.267 |
| blendface | 2023 | FS | 0.943 | 0.081 | 0.040 |
| facedancer | 2023 | FS | 0.933 | 0.070 | 0.012 |
| inswap | 2023 | FS | 0.815 | 0.158 | 0.074 |
| pixart | 2023 | EFS | 0.433 | 0.580 | 0.514 |
| sadtalker | 2023 | FR | 0.427 | 0.555 | 0.477 |
| SiT | 2024 | EFS | 0.554 | 0.436 | 0.402 |
| rddm | 2024 | EFS | 0.733 | 0.150 | 0.303 |

**Table 9.** *Label-free signals against competence on the 21-generator timeline set (Xception). Each signal is computed without labels; AUC uses labels, making this a non-circular test. Reference-divergence and distribution-shape signals carry genuine competence information here; their cross-architecture portability, where the reference-free signals prove more robust than reference-divergence, is quantified in the text. The mean score (marked) is the strongest correlate but is class-balance-confounded and not used as a deployable signal.*

| Signal | r vs AUC | p | Spearman |
|---|---|---|---|
| mean_score (excluded) | 0.993 | 2.7e-19 | 0.991 |
| ks_vs_ref | 0.952 | 3.3e-11 | 0.948 |
| wasserstein_vs_ref | 0.884 | 1.1e-7 | 0.884 |
| bimodality | 0.824 | 4.5e-6 | 0.804 |
| std_score | −0.739 | 1.3e-4 | −0.764 |
| entropy | −0.681 | 6.8e-4 | −0.736 |
| confident_frac | 0.567 | 0.007 | 0.756 |
| kl_vs_ref | 0.265 | 0.245 | 0.764 |

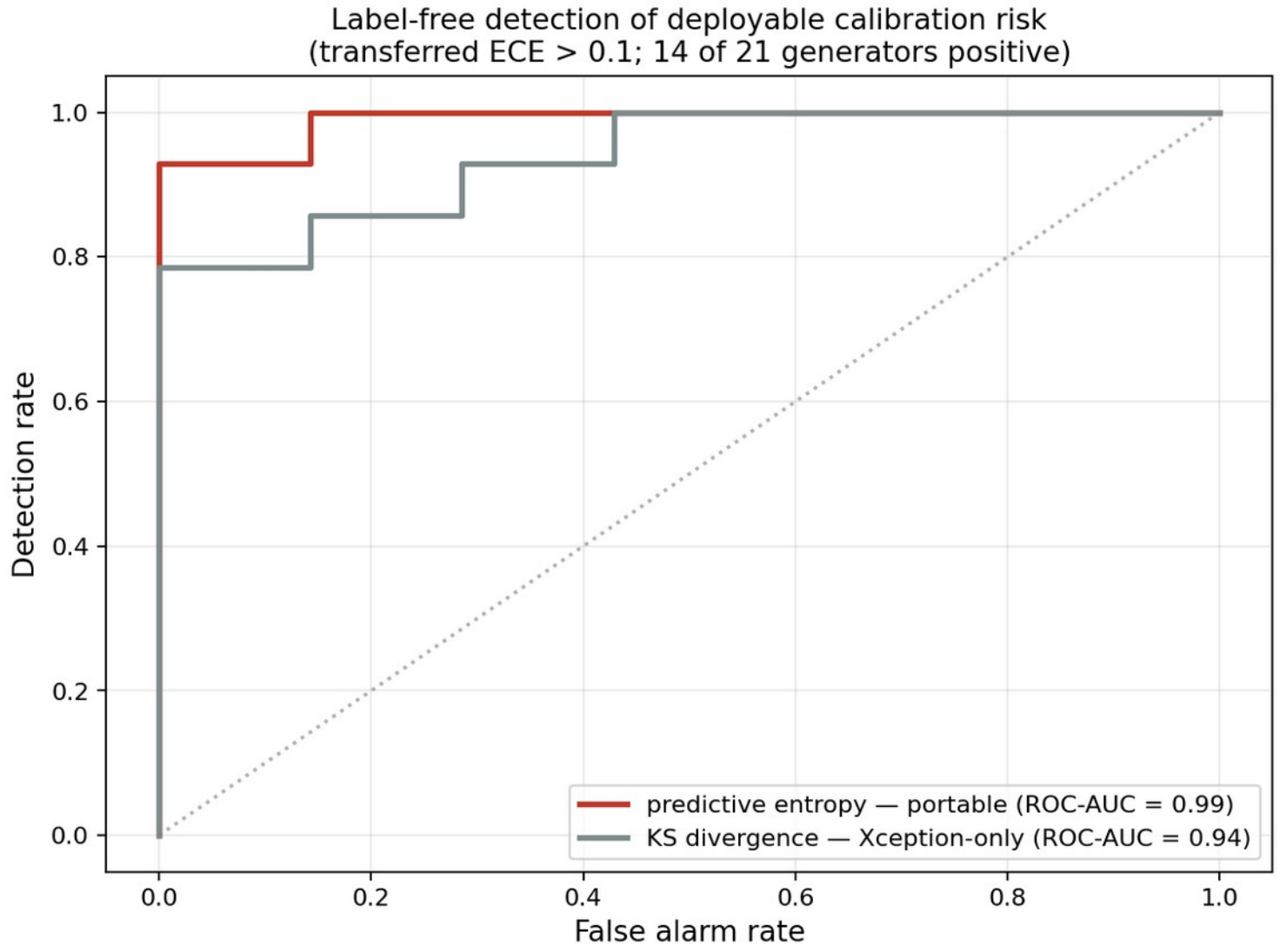

**Figure 9.** *Label-free detection of deployable calibration risk (transferred ECE > 0.1; 14 of 21 generators positive). Predictive entropy, the portable reference-free monitor, flags the at-risk generators at ROC-AUC 0.99 (95% CI [0.94, 1.00]); KS divergence scores 0.94 here but its orientation is Xception-specific and reverses on other backbones (Table 10), so entropy is the signal relied on for portability.*

**Table 10.** *Cross-architecture behavior of label-free monitoring signals: Pearson correlation of each signal with competence (AUC) on each detector. The key pattern is directional: predictive entropy keeps a consistent negative sign on all three backbones, whereas the reference-divergence signals (KS, Wasserstein) actively reverse between Xception (strongly positive) and EfficientNet (negative), making a threshold tuned on one backbone point the wrong way on another. Score dispersion is reliable only where the score distribution has range: it is strongly negative on Xception and the transformer but compresses to a narrow band on EfficientNet (standard deviation 0.29 to 0.37 across generators) where its residual correlation is weakly positive (+0.53), so dispersion is not a dependable cross-architecture monitor and entropy is the signal we rely on. EfficientNet is the hard case for all distribution-shape signals: entropy weakens to −0.38 there but holds its sign, while dispersion loses its range and the reference-divergence signals invert outright. All correlations are over the 20 DF40 generators except the Xception dispersion row (21-generator timeline set). CLIP reference-divergence is computed against an in-domain DF40 face-swap reference pool (no CLIP-scored FaceForensics++ reference exists); the CNN columns use the FaceForensics++ faceswap-and-real reference.*

| Signal | Family | Xception (CNN) | EfficientNet (CNN) | CLIP (ViT) |
|---|---|---|---|---|
| predictive entropy | reference-free | −0.84 | −0.38 | −0.95 |
| score dispersion | reference-free | −0.73 | +0.53 | −0.96 |
| KS vs reference | reference-divergence | +0.96 | −0.34 | −0.94 |
| Wasserstein vs reference | reference-divergence | +0.90 | −0.09 | −0.97 |

## 8 Competence as a Shared Driver of the Deployable Trust Signals

Sections 4 through 7 each relate a trust signal to competence. We now ask whether these are facets of one structure, on the quantity a deployment actually consumes: the transferred trust score. On the 20-generator unifying set, transferred ECE correlates with competence at $r = -0.99$, with explanation faithfulness at $r = -0.91$, and with the monitoring signals at $|r|$ between 0.83 and 0.92 (Figures 10–11). Two cautions from the preliminary version are now made precise. First, we no longer present the two-variable principal component of calibration and faithfulness as evidence of a latent factor: for two standardized variables the first component explains $(1 + |r|)/2$ of the variance by identity, so that analysis restates the correlation, and we report the correlation. Second, attenuation under adjustment is now tested formally: bootstrapping the paired difference $\Delta = |r_raw| - |r_partial|$ over generators, conditioning on competence significantly attenuates every pairwise association among transferred ECE, faithfulness, and the monitoring signal ($\Delta$ 95% CIs [+0.20, +0.92], [+0.06, +0.79], and [+0.15, +0.91], all excluding zero). This is evidence of mediation-consistent structure, not proof of a latent causal factor: the associations are generator-level at n = 20, the five-signal decomposition remains exploratory, and only the calibration link carries the induced-competence manipulation of Section 4.4. With those bounds stated, the synthesis is unchanged in substance and stronger in measurement: on the deployable quantities, the trust signals degrade together, and the axis they share is competence.

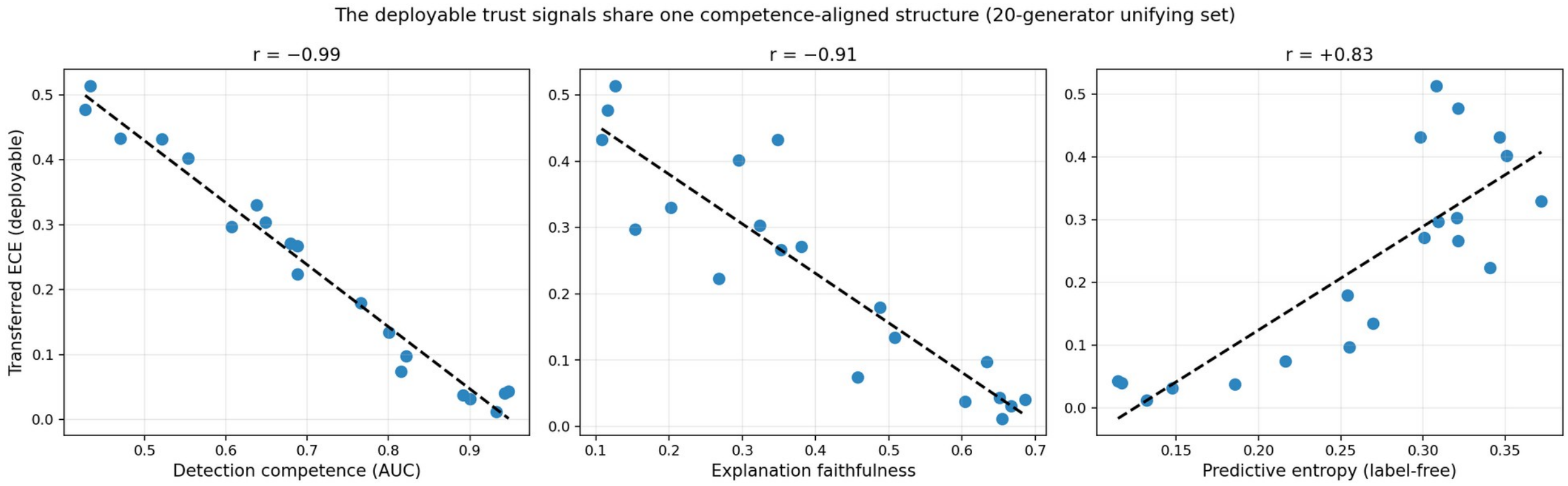

**Figure 10.** *The deployable trust signals share one competence-aligned structure. Transferred ECE against competence, explanation faithfulness, and label-free predictive entropy on the 20-generator unifying set.*

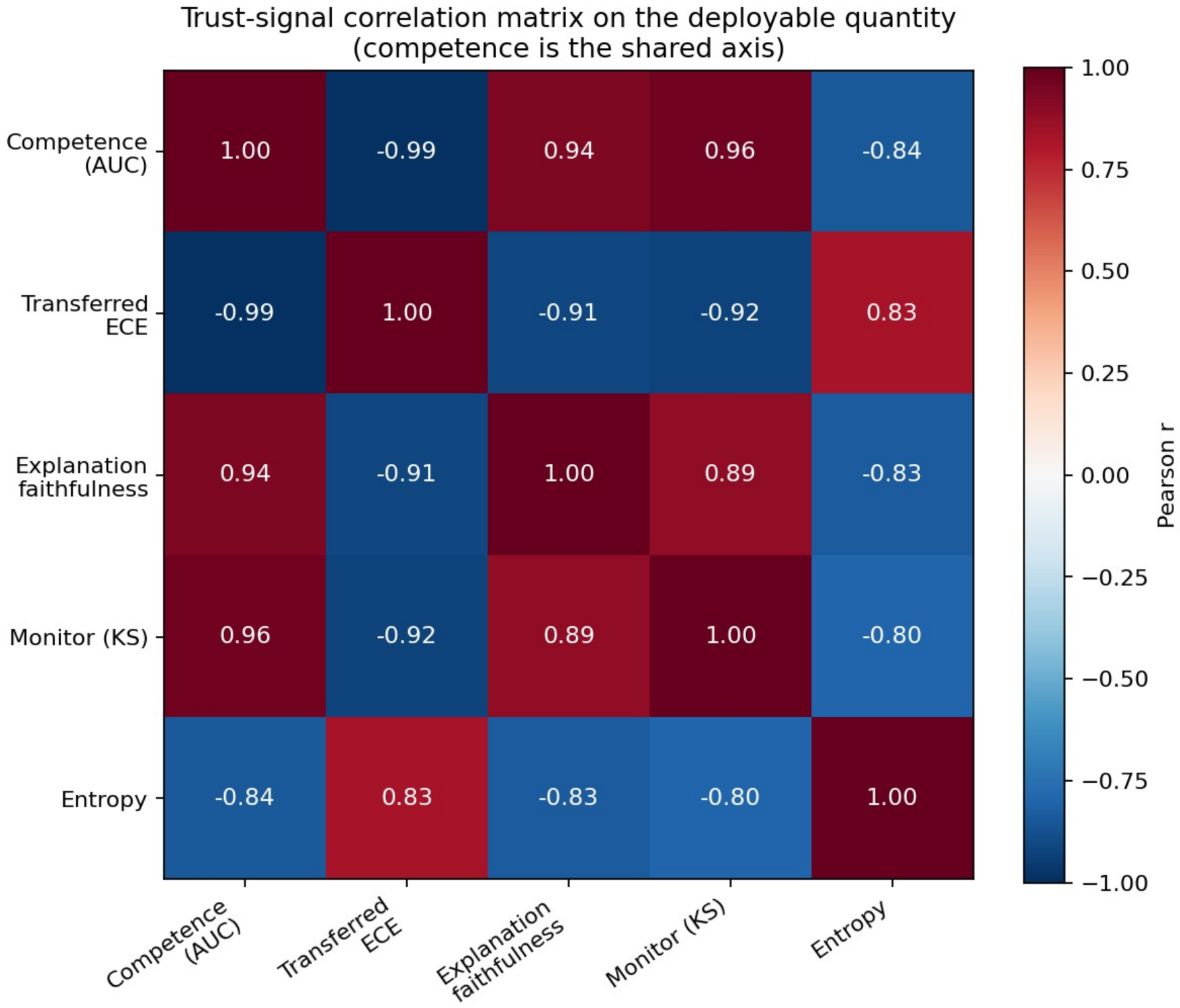

**Figure 11.** *Trust-signal correlation matrix on the deployable quantity. Every signal co-varies with every other; competence (AUC) is the shared axis.*

## 9 Competence-Aware Routing

If competence governs every trust failure and is trackable without labels, then a verification policy should abstain on low competence rather than low confidence. This matters because the two come apart exactly where it counts: a low-competence detector is confidently wrong, so confidence is the wrong signal to route on.

## 9.1 Setup

We cast routing as selective prediction (El-Yaniv & Wiener, 2010; Geifman & El-Yaniv, 2017). At coverage c we answer the c fraction of inputs a routing signal deems most trustworthy and abstain on the rest, measuring selective error on the answered set and summarizing each signal by the area under the risk-coverage curve (AURC; lower is better). The pool for each detector is its full DF40 suite: 593,502 frames across 21 generators for Xception, and 573,855 frames across the 20-generator unifying set for EfficientNet-B4 and the CLIP transformer, so routing is evaluated on all three architectures. We compare per-frame confidence, per-batch score dispersion, per-batch mean predictive entropy, and, as a comparison row, per-batch KS divergence from the in-domain reference; the batch is the source-batch (here, the generator), the granularity at which deployed content arrives from a given uploader, campaign, or generator. Because a batch-level signal ties every frame within a batch, the within-batch order is randomized and every reported number is the mean over ten random orders (the seed spread is below $10^{-3}$ throughout); each batch signal is routed in the direction its competence coupling on that backbone prescribes (Section 7.2, Table 10). Per-frame predictive entropy is a monotone function of $|p - 0.5|$ and is therefore mathematically equivalent to confidence, so it is not reported separately.

## 9.2 Competence-aware routing wins exactly where the coupling says confidence fails

Table 11 reports the Xception result and Table 12 the cross-architecture picture; Figures 12 and 13 show the risk-coverage curves. On Xception, routing on batch dispersion beats per-frame confidence overall (AURC 0.212 versus 0.221), with the advantage concentrated in the low-to-mid coverage regime a verification gate occupies (selective error 0.153 versus 0.206 at 40% coverage, 0.281 versus 0.287 at 70%) and a small crossover at high coverage (0.310 versus 0.303 at 80%), where almost nothing is abstained. On EfficientNet-B4, the harder, lower-competence pool (full-coverage error 0.470), both reference-free batch signals beat confidence, batch entropy most (AURC 0.354, dispersion 0.373, confidence 0.385). On the CLIP transformer the ordering reverses: per-frame confidence wins outright (AURC 0.106 against 0.157 for dispersion), and the reversal is the coupling's own prediction rather than a failure of the method. The CLIP pool is high-competence almost everywhere (full-coverage error 0.252), and precisely where competence is high the coupling says the calibrated score is trustworthy, so a reliable per-frame signal outranks any batch-level surrogate that discards within-batch information. The regime in which each routing signal wins is therefore itself organized by competence: batch-level competence routing pays on the low-competence pools where confidence is confidently wrong (both CNNs), and per-frame confidence pays where competence is high (the transformer).

Two further results sharpen the portability analysis of Section 7.2 operationally. First, the KS comparison row: on Xception, batch KS attains the best in-sample AURC (0.194), but on EfficientNet, where its orientation reverses (Table 10), KS routing collapses to AURC 0.532, worse than answering every frame with no abstention at all (0.470), and on CLIP it reaches only 0.356 against 0.106 for confidence. A reference-divergence router tuned on one backbone does not merely underperform on another; it actively routes the wrong batches. Second, the direction problem: dispersion required its backbone-specific orientation from Section 7.2 (answered low-dispersion-first on Xception and CLIP but high-dispersion-first on EfficientNet, where dispersion couples positively with competence), whereas batch entropy keeps one direction on all three backbones. Entropy is therefore the only batch signal whose routing direction is knowable a priori on a new backbone, the same portability conclusion the monitoring analysis reached; dispersion is the stronger ranker where its orientation is known, and reference-divergence should not be used for routing at all.

**Table 11.** *Risk-coverage on Xception: AURC (lower is better) and selective error at fixed coverage, pooled over 21 generators (593,502 frames). Within-batch order is randomized; every value is the mean over ten random orders (seed spread below $10^{-3}$). Batch dispersion achieves a lower overall AURC than per-frame confidence with its advantage at low-to-mid coverage; batch entropy does not improve on confidence on this backbone; batch KS is the best in-sample but is not architecture-portable (Table 12).*

| Routing signal | AURC | Err@40% | Err@70% | Err@80% | Err@90% |
|---|---|---|---|---|---|
| confidence (per-frame) | 0.221 | 0.206 | 0.287 | 0.303 | 0.317 |
| batch dispersion (ref-free) | 0.212 | 0.153 | 0.281 | 0.310 | 0.326 |
| batch entropy (ref-free) | 0.222 | 0.170 | 0.308 | 0.313 | 0.336 |
| batch KS (ref-divergence) | 0.194 | 0.156 | 0.239 | 0.279 | 0.306 |

**Table 12.** *Cross-architecture routing: AURC (lower is better) for each signal on each detector's full DF40 pool, within-batch order randomized and averaged over ten seeds. Batch competence routing wins on the two lower-competence CNN pools; per-frame confidence wins on the high-competence transformer pool, the regime dependence the coupling predicts. KS routing on EfficientNet is worse than no abstention (0.532 > 0.470), the operational cost of the Table 10 sign reversal. Dispersion is routed in its backbone-specific direction from Table 10; entropy keeps one direction on all three backbones. The CLIP KS reference is the in-domain DF40 face-swap pool (no CLIP-scored FaceForensics++ reference exists).*

| Architecture (pool) | No-abstention error | Confidence | Batch dispersion | Batch entropy | Batch KS |
|---|---|---|---|---|---|
| Xception (21 gen; 593,502 frames) | 0.334 | 0.221 | 0.212 | 0.222 | 0.194 |
| EfficientNet-B4 (20 gen; 573,855) | 0.470 | 0.385 | 0.373 | 0.354 | 0.532 |
| CLIP-ViT (20 gen; 573,855) | 0.252 | 0.106 | 0.157 | 0.159 | 0.356 |

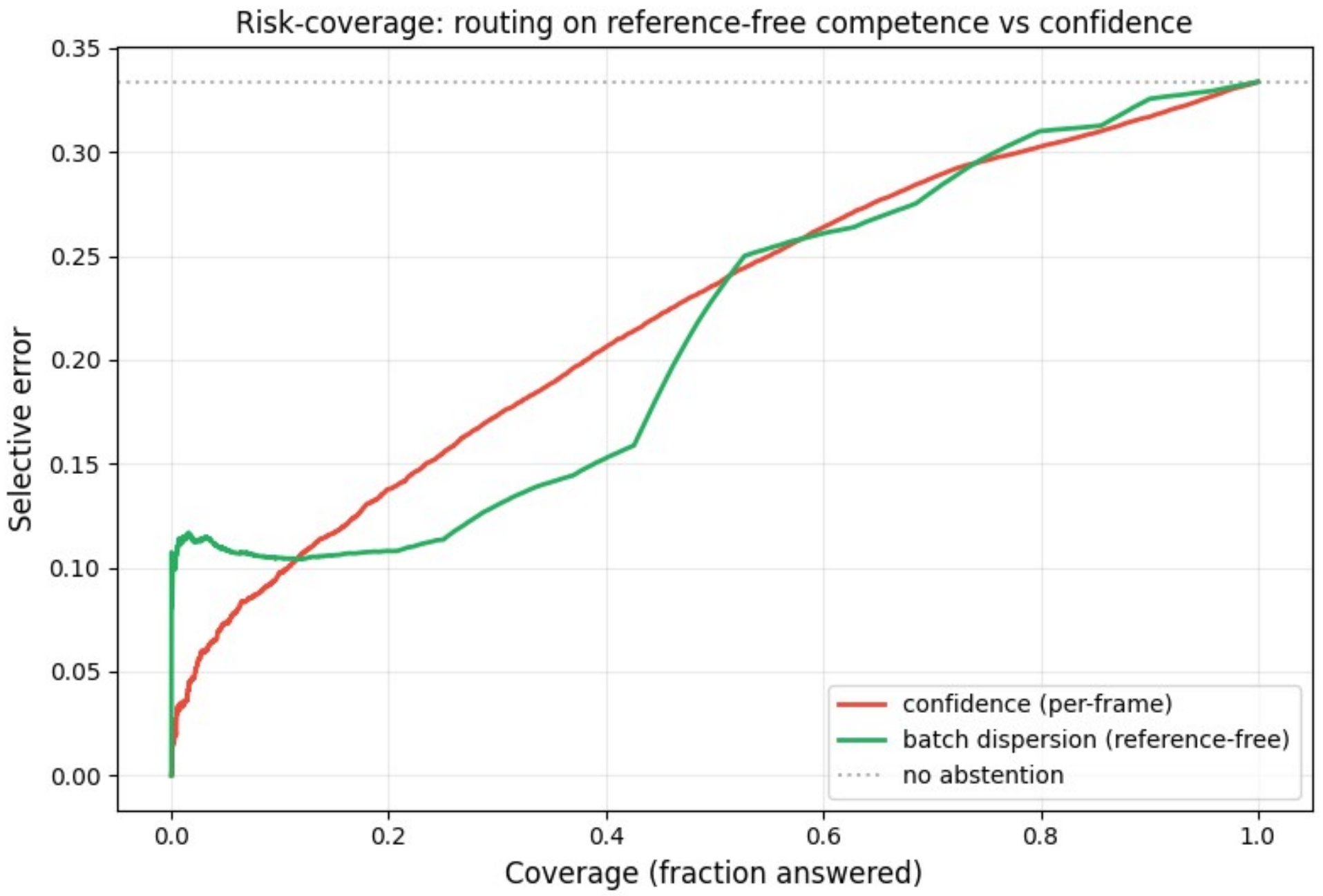

**Figure 12.** *Risk-coverage on Xception. Routing on batch-level reference-free competence (score dispersion, green) achieves a lower overall risk-coverage curve than per-frame confidence (red), with the advantage concentrated at low-to-mid coverage and a small crossover at high coverage. Within-batch order randomized (seed 0 shown); Table 11 reports ten-seed means.*

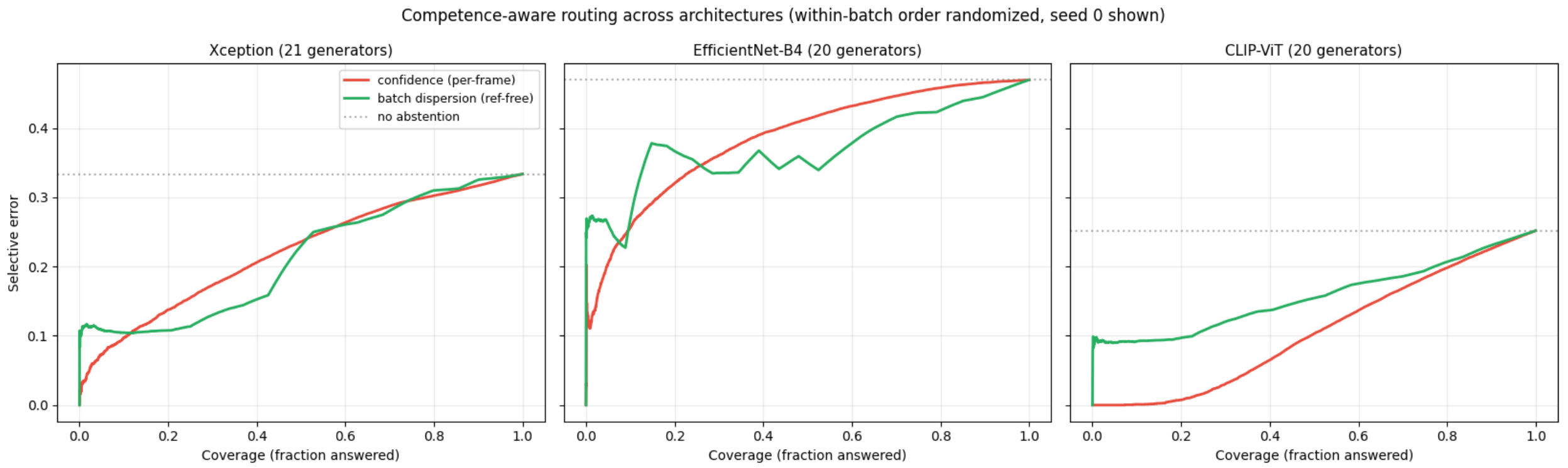

**Figure 13.** *Risk-coverage across architectures. Batch-dispersion routing (green) beats per-frame confidence (red) on the two lower-competence CNN pools (left, centre) and loses to it on the high-competence transformer pool (right), the regime dependence the coupling predicts. Within-batch order randomized; seed 0 shown.*

# 10 Discussion

## 10.1 One factor, not five guarantees

The engineering upshot survives the correction and is cleaner for it. What a deployment must fear is not that its detector cannot be calibrated, it can, cheaply, wherever labels exist, but that its calibration silently stops transferring exactly where competence drops, that its raw score is overconfident in proportion to its incompetence, and that its explanations lose faithfulness on the same axis. All of these are governed by competence; competence is lowest on the novel generators that drive real risk; and a batch's competence can be tracked without labels accurately enough to flag its deployed calibration failures (Section 7). The operating recipe is unchanged: monitor competence and condition trust. What the correction removes is a false comfort and a false alarm at once: the false alarm that low-competence scores are beyond calibration even with labels, and the false comfort that a fixed calibration certificate means anything off-distribution.

## 10.2 Deployment implications

Three common practices fail for the same reason. Static calibration on a fixed benchmark, a fairness audit on a fixed dataset, and confidence-threshold routing all ignore competence, which is lowest precisely for the novel generators that drive real risk (Chandra et al., 2025). CDTS conditions on competence end to end: calibrate in-domain and treat the certificate as distribution-bound (Section 4), audit calibration equity with a tie-safe estimator, treating a null as informative (Section 5), estimate competence label-free from the score distribution on generators where labels do not (Section 7), and route source-batches on that label-free estimate (Section 9). The operating recipe is competence-centered: monitor competence and condition trust.

## 10.3 Limitations

Several limitations bound these claims. The equity audit is single-detector at 70 identities and is now a bounded null rather than a positive finding. The explanation-faithfulness and full multi-signal analyses are conducted on the two convolutional detectors, and the mediation evidence is associational except for the calibration link, which carries the induced-competence manipulation. The faithfulness metric validates explanations against the model's own output behaviour; we do not additionally run parameter-randomization sanity checks (Adebayo et al., 2018), so faithfulness claims are relative to that output-fidelity criterion. The transferred-calibrator analysis is one backbone (Xception) and one in-domain pool; the cross-backbone transferred analysis is future work. The CLIP reference-divergence comparison mixes an architecture change with a reference change (Section 7.2). Configuration dependence is addressed by family-cluster and leave-one-family-out analyses but not by a full hierarchical model, and both AUC and ECE are estimated quantities, negligible at these split sizes. The label-free monitor is an information source, not yet a validated estimator with prediction intervals. The routing advantage is realized at the batch level and which signal wins depends on the pool's competence regime (Section 9.2). Finally, the estimator correction itself: we validated it against an exact conditional estimator and a committed regression test, and we retain the legacy estimator so every preliminary number remains reproducible, but the community should scrutinise it; algorithmic recourse (Mothilal et al., 2020) remains future work.

## 11 Conclusion

Deepfake-detector trustworthiness is organized, to a substantial degree, by one measurable factor: detection competence. We showed that the raw trust score's miscalibration tracks competence at $r = -0.98$ across two convolutional networks and a vision transformer; that on the primary backbone a frozen calibrator, deployed without target labels, fails with competence at the same near-perfect strength for every calibrator family we tested; that, given labels, calibration is cheap at any competence, so the failure is a distribution-shift phenomenon rather than a calibratability one; that explanation faithfulness rises and falls on the same axis; and that competence is trackable without labels well enough to flag deployable calibration failures at ROC-AUC 0.99 and to route verification traffic where confidence cannot be trusted. We also showed something we did not set out to find: that the standard equal-mass ECE estimator harbours a tie-handling degeneracy that manufactures competence–calibration coupling and calibration-inequity where none is measurable, an artifact that shaped the preliminary version of this work and plausibly others. Correcting it made the paper's deployable claims stronger and its in-domain claims honest. Because competence is lowest exactly where deployment risk is highest, trust scoring must be competence-aware; the CDTS wrapper, know, show, warn, and route, all conditioned on competence, is the mechanism, and its measurement now stands on an estimator that tells the truth at the extremes.

#### CRediT authorship contribution statement

Md Anas Biswas: Conceptualization, Methodology, Software, Validation, Formal analysis, Investigation, Data curation, Writing - original draft, Writing - review and editing, Visualization, Project administration.

#### Declaration of competing interest

The author declares no competing financial interests or personal relationships that could have appeared to influence the work reported in this paper.

### Declaration of generative AI in the writing process

During the preparation of this work the author used Claude (Anthropic) to assist with drafting, editing, statistical verification, and internal consistency auditing of the manuscript, under the author's direction and using the author's data and analyses. The author reviewed, verified, and edited all content and takes full responsibility for the content of this article.

#### Funding

This research did not receive any specific grant from funding agencies in the public, commercial, or not-for-profit sectors.

#### Data availability

The analysis code, per-configuration results, per-frame detector scores, and audit scripts are available at https://github.com/anasbiswas1/deepfake-trust-research. FaceForensics++, Celeb-DF v2, DF40, the DeepFakeDetection set, and the demographic annotations of Xu et al. (2024) are third-party datasets available from their original providers under their respective licenses (DF40 under CC BY-NC 4.0); they are not redistributed.

## Appendix A Architecture Robustness of the Unifying Result: Pre-Specified Test and Outcome

Preservation note. This appendix records a pre-specified test and its outcome exactly as originally computed and reported. Its ECE inputs used the legacy estimator, whose tie-handling degeneracy is documented in Section 3.5, so the specific variance shares and correlations below are superseded by the corrected analyses of Sections 4 and 8, and for the intrinsic-factor claim the corrected quantities reverse the verdict (see the postscript); the numbers are retained solely as a pre-registration record; the two-variable variance shares additionally restate correlations by the $(1 + |r|)/2$ identity noted in Section 8. The appendix's methodological point, that pre-specifying the cross-architecture test and reporting its partial failure bounds the claim honestly, stands.

We pre-specified the following test of the unifying result before computing it: repeat the full multi-signal principal-component analysis on EfficientNet-B4 (Tan & Le, 2019), and require the first component to explain at least 75% of the variance and to align with competence at $|r|$ of at least 0.90 for the single-factor structure to count as architecture-general; a materially weaker or non-aligned component would bound the full-signal claim to Xception. We report the outcome honestly. **The full-signal bar was not met on EfficientNet.** The EfficientNet full five-signal first component explains only 44.0% of the variance (against an Xception value of 84.7%), with a second component of nearly equal size, and aligns with competence at $r = 0.82$. Both fall short of the pre-specified thresholds. Diagnosing why (Section 7.2) shows the cause: the reference-divergence monitoring signals decouple from competence on EfficientNet and form a second axis, splitting the variance. **What did replicate.** Restricting to the intrinsic trust signals (calibration and explanation faithfulness) the single-factor structure holds on both convolutional architectures: the intrinsic first component explains 90.7% of the variance on Xception ($r = 0.95$ with competence) and 86.8% on EfficientNet ($r = 0.85$). We additionally tested a third, non-convolutional architecture, a CLIP vision transformer: the competence-calibration coupling holds there at $r = -0.859$ (the legacy value; Table 4 reports the corrected grid), and reference-free uncertainty signals track competence on the transformer at up to $r = -0.962$ (Section 7.2). The transformer has no convolutional feature map and so cannot contribute a Grad-CAM explanation-faithfulness signal, which is why it does not enter the intrinsic-factor decomposition. **Conclusion of the test.** The pre-specified full-signal single factor is architecture-specific to Xception. What the paper claims after the estimator correction is stated in Section 2.5 and Section 8: the raw and transferred competence–trust couplings (three architectures, CNN and transformer), reference-free label-free monitoring (three architectures), and the faithfulness coupling (both convolutional detectors). The pre-specified full-signal single factor is not among the claims, and neither, after the correction, is the intrinsic two-signal factor reported above. The value this appendix retains is methodological: the acceptance criteria were fixed before the result was known, and the outcome, a partial failure then and a fuller one under corrected measurement, is reported rather than repaired.

Postscript under the corrected estimator. Re-examining this test's ingredients with the tie-safe estimator of Section 3.5 reverses the surviving claim rather than merely blurring it. The intrinsic two-signal factor rested on a strong correlation between target-fitted calibration error and explanation faithfulness; under the corrected estimator that correlation is $|r| = 0.12$ on Xception and 0.02 on EfficientNet-B4, so the intrinsic factor was a product of the estimator degeneracy and the two-variable variance identity, and we withdraw it. What carries competence-aligned structure under correct measurement is the deployable quantity: on the unifying set, transferred ECE correlates with explanation faithfulness at $|r| = 0.91$ and with competence at 0.99, the analysis reported in Section 8. The pre-specified test's architecture-scoping instinct was right; its passing grade on the primary architecture was not, because one of its five signals was measured with the degenerate estimator.

## Appendix B  Configuration Sets and Where Each Is Used

The analyses draw on several overlapping configuration sets. The inventory below states each set, its size, its contents, and the sections and exhibits that use it. The per-detector raw-coupling coefficients (r = −0.982, −0.983, −0.950) are computed on the three full detector suites, whose union is the 61-configuration pool of Section 4; the transferred coupling is computed on the 21-generator Xception timeline; and the four FR-mismatch configurations are a distinct induced set that belongs to no suite.

| Configuration set | n | Contents | Where used |
|---|---|---|---|
| Xception suite | 21 | FS-Xception on all DF40 timeline generators (rddm included) | Tables 2, 4, 5; Figures 3–5; transferred setting; Section 9 pool |
| EfficientNet-B4 suite | 20 | Unifying set scored by EfficientNet-B4 | Tables 2, 4; Figure 3 |
| CLIP-ViT suite | 20 | Unifying set scored by CLIP-ViT | Tables 2, 4; Figure 3 |
| 61-configuration pool | 61 | Union of the three suites (AUC 0.43–0.99) | Sections 4.1–4.2, 4.7; Figures 3–4; Tables 2, 5 |
| FR-mismatch set | 4 | Reenactment-trained Xception on simswap, blendface, StyleGAN2, sd2.1 (AUC 0.26–0.46) | Sections 4.2, 4.4; Figure 2; Appendix C |
| Within-DF40 confound set | 12 | Eight FS-suite configurations plus the four FR-mismatch | Table 2; Figure 2 |
| FF++ slices | 12 | FS and FR Xception (8) and FS EfficientNet (4) per-method configurations | Table 2 (grand pool); Appendix C (FR) |
| Grand raw pool | 77 | All raw slices above combined | Table 2 |
| 21-generator timeline | 21 | Identical to the Xception suite; era labels attached | Tables 8–9; Figure 9; Section 7 |
| 20-generator unifying set | 20 | Timeline minus rddm; scored on all three detectors | Sections 7–8; Table 10; Figures 10–11 |

## Appendix C  Orientation and Calibrator Audit of the Low-Competence Configurations

For every induced-low-competence configuration, every sub-0.5 EfficientNet configuration, and two high-competence controls, the table reports native AUC, orientation-free AUC* = max(AUC, 1 − AUC), the fitted Platt slope, and evaluation-half ECE as follows. The Hybrid column is the preliminary calibrator under the legacy estimator, reproduced for contrast; every other calibrated column, orientation-selected isotonic (direction chosen on the calibration split only), unconstrained Platt, and beta calibration, uses the corrected tie-safe estimator of Section 3.5 (for Platt and beta the two estimators coincide to the reported precision, since their outputs are continuous and tie-free; we verified the agreement exactly). The final column reports the constant-prevalence predictor under both estimators, replicating the Section 3.5 demonstration on every configuration: tie-safe charges it the prevalence shift (0.012–0.035) while the legacy estimator charges 0.27–0.47, and where the Hybrid column is large it coincides with the legacy constant-prevalence value, the signature of a collapsed calibrator scored by the degenerate estimator rather than of genuine miscalibration. The Platt slope is negative on all four genuinely inverted induced configurations and near zero of either sign on the no-signal EfficientNet configurations. Well-specified calibrators achieve low error at every competence. All splits are identity-disjoint (seed 42), matching the main protocol.

| Configuration | AUC | AUC* | Platt slope | Hybrid (legacy est.) | Iso (oriented) | Platt | Beta | Const-prev (tie-safe / legacy) |
|---|---|---|---|---|---|---|---|---|
| XcpFR-DF40 StyleGAN2 | 0.256 | 0.744 | −0.17 | 0.268 | 0.018 | 0.041 | 0.032 | 0.012 / 0.268 |
| XcpFR-DF40 blendface | 0.457 | 0.543 | −0.02 | 0.248 | 0.024 | 0.066 | 0.021 | 0.022 / 0.324 |
| XcpFR-DF40 sd2.1 | 0.329 | 0.671 | −0.10 | 0.268 | 0.014 | 0.027 | 0.022 | 0.012 / 0.268 |
| XcpFR-DF40 simswap | 0.363 | 0.637 | −0.07 | 0.301 | 0.033 | 0.060 | 0.031 | 0.035 / 0.327 |
| XcpFR-FF++ faceswap | 0.387 | 0.613 | −0.10 | 0.467 | 0.060 | 0.059 | 0.057 | 0.015 / 0.467 |
| XcpFR-FF++ face2face | 0.495 | 0.505 | +0.01 | 0.341 | 0.041 | 0.032 | 0.031 | 0.016 / 0.467 |
| XcpFR-FF++ neuraltextures | 0.640 | 0.640 | +0.12 | 0.158 | 0.055 | 0.039 | 0.042 | 0.015 / 0.467 |
| XcpFR-FF++ deepfakes | 0.664 | 0.664 | +0.10 | 0.169 | 0.061 | 0.045 | 0.057 | 0.017 / 0.467 |
| EffNet-DF40 StyleGAN2 | 0.466 | 0.534 | +0.01 | 0.181 | 0.018 | 0.020 | 0.013 | 0.012 / 0.268 |
| EffNet-DF40 StyleGAN3 | 0.477 | 0.523 | +0.01 | 0.206 | 0.020 | 0.020 | 0.016 | 0.012 / 0.268 |
| EffNet-DF40 StyleGANXL | 0.477 | 0.523 | +0.04 | 0.191 | 0.016 | 0.028 | 0.014 | 0.012 / 0.268 |
| EffNet-DF40 pixart | 0.472 | 0.528 | +0.00 | 0.205 | 0.015 | 0.049 | 0.012 | 0.012 / 0.268 |
| EffNet-DF40 sadtalker | 0.489 | 0.511 | +0.01 | 0.254 | 0.021 | 0.051 | 0.022 | 0.018 / 0.330 |
| CTRL XcpFS simswap | 0.948 | 0.948 | +0.50 | 0.024 | 0.019 | 0.023 | 0.020 | 0.035 / 0.327 |
| CTRL XcpFS wav2lip | 0.470 | 0.530 | +0.00 | 0.224 | 0.025 | 0.045 | 0.018 | 0.017 / 0.316 |